\documentclass[11pt]{article}
\usepackage{amssymb,amsmath,amsfonts}
\usepackage{graphicx}
\usepackage{graphics}
\usepackage{eepic,epsfig}

\begin{document}

\title{Finite temperature energy-momentum tensor in compactified cosmic string spacetime}
\author{W. Oliveira dos Santos$^{1}$%
\thanks{%
		E-mail: wagner.physics@gmail.com} ,\thinspace\ E. R. Bezerra de Mello$^{1}$\thanks{%
		E-mail: emello@fisica.ufpb.br} \\
	\\
	$^{1}$\textit{Departamento de F\'{\i}sica, Universidade Federal da Para\'{\i}%
		ba}\\
	\textit{58.059-970, Caixa Postal 5.008, Jo\~{a}o Pessoa, PB, Brazil}}
\maketitle

\begin{abstract}
In this paper we analyze the expectation value of the field squared and the energy-momentum tensor associated with a massive charged scalar quantum field with a nonzero chemical potential propagating in a high-dimensional compactified cosmic string spacetime in thermal equilibrium at finite temperature $T$. Moreover, we assume that the charged quantum field interacts with a very thin magnetic flux running along the core of the idealized cosmic string, and with a magnetic flux enclosed by the compact dimension. These observables are expressed as the vacuum expectation values and the finite temperature contributions coming from the particles and antiparticles excitations. Due to the compactification, the thermal corrections can be decomposed in a part induced by the cosmic string spacetime without compactification, plus a contribution induced by the compactification. This decompositions explicitly follows from the Abel-Plana formula used to proceed the summation over the discrete quantum number associated with the quasiperiodic condition imposed on the quantum field along the compact dimension. The expectations values of the field squared and the energy-momentum tensor are even periodic functions of the magnetic flux with period being the quantum flux, and also even functions of the chemical potential. Our main objectives in this paper concern in the investigation of the thermal corrections only. In this way we explicitly calculate the behavior of these observables in the limits of low  and  high temperature. We show that the temperature enhance the induced densities. In addition some graphs are also included in order to exhibit these behaviors. 
\end{abstract}

\bigskip


\bigskip

\section{Introduction}
According to Big Bang theory, at the beginning the Universe was hotter and in a more symmetric stage. During its expansion
process it has cooled down and underwent to a series of phase transitions accompanied by spontaneous breakdown of symmetries which could result in the formation of topological defects \cite{Kibble,V-S}. These include domain walls, cosmic strings and monopoles. Among them the cosmic strings are of special interest.
 
 Cosmic strings are lines of trapped energy density, analogous to defects such as vortex lines
 in superconductors and superfluids. These object modifies the topology of the spacetime,  and can be of cosmological
 and astronomical significance in a large number of phenomena, such as producing cosmic microwave background anisotropies, non-Gaussianity and B-mode polarization, sourcing gravitational waves, generation of high energy cosmic rays and gravitationally lensing astrophysical objects \cite{Hindmarsh2, Kawasaki2010, Virbhadra2009, Fraisse2008, Pogosian2008, Virbhadra2000, Laix1997, Kaiser1984, Vilenkin1981}. The dimensionless parameter that characterizes the strength of gravitational interactions of strings with matter is its {\it tension}, that is given in natural units by $G\mu_0$, being $G$ the Newton's constant and $\mu_0$ its linear mass density, proportional to the square of the symmetry breaking scale energy.

The gravitational field produced by an idealized cosmic string may be approximated by a planar angle deficit in the two-dimensional sub-space orthogonal to the string.  Although there is no Newtonian potential, the lack of global flatness is responsible for many interesting phenomena as shown many years ago by Linet \cite{Linet} and Smith \cite{Smith}. 
Moreover, the presence of  the string allows effects such as  particle-antiparticle pair production by a single photon and bremsstrahlung radiation from charged particles which are not possible in empty Minkowski space due to the conservation
of linear momentum \cite{Skarzhinsky}. 

Another type of topological quantum effects considered in the literature is induced by compact spatial dimensions. The presence of compact dimensions is a important feature of most high energy theories of fundamental physics, including
supergravity and superstring theories. An interesting application of the field theoretical models with compact dimensions recently appeared in nanophysics. The long-wavelength description of the electronic states in
graphene can be formulated in terms of the Dirac-like theory in three-dimensional spacetime with the Fermi velocity playing the role of speed of light (see, e.g., \cite{Cast09}).

The Casimir effects caused by the compactification on bosonic vacuum in  $(1+D)$-dimensional cosmic string spacetime  was first analyzed in \cite{Saha_2012}. There a quasiperiodic condition on the field along the compact dimension has been considered. The analysis of the induced current, associated with a massive charged scalar quantum field in $(1+D)$-dimensional compactified cosmic tring spacetime, has been developed in \cite{Braganca_15}, assuming the presence of a magnetic flux running along the string's core, and with an additional magnetic flux enclosed by the compact dimension. Moreover, considering still this system, the investigations of the vacuum expectation values of field squared and energy-momentum tensor have been carried out in \cite{Braganca_19}. Continuing in this line of investigations, in the present paper we will consider the effects of the finite temperature and nonzero chemical potential on the expectation values on these observables.  This is an important topic, since for a cosmic string in the early stages of the cosmological expansion of the Universe,  the typical state of a quantum field is a state containing particles and anti-particles in thermal equilibrium at finite temperature $T$. The finite temperature	expectation value of the energy density for a massless scalar field around a cosmic string in the absence of the magnetic flux is derived in \cite{Davi88}	for integer values of the parameter $q=2\pi /\phi _{0}$, where $2\pi -\phi_{0}$ is the planar angle deficit caused by the presence of the cosmic string. The expectation value of a renormalized energy-momentum tensor for a general case of the parameter $q$	has been considered in \cite{Line92} for a conformally coupled massless scalar field and in \cite{Frol95} for a general case of a curvature coupling parameter. Guimar\~{a}es \cite{Guim95} has extended the corresponding results considering the presence of a magnetic flux running along the  cosmic string assuming that $q<2$. In these last four papers,  the thermal Green functions have been obtained by imposing periodic condition on the Euclidean time with period $\beta=1/T$ in the corresponding zero-temperature Green functions.

The thermal correction to the current density previously investigated in \cite{Braganca_15} was calculated in \cite{Mohammadi_16}. In the latter it was shown that the induced charge density is an odd function of chemical potential, $\tilde{\mu}$. So when this parameter is zero, the contribution of particle and antiparticle cancel each other. The chemical potential imbalances these contributions. As to the azimuthal current density it is a even function of $\tilde{\mu}$.

This paper is organized as follows: In section \ref{sec2} we introduce the model setup that we want to study, and presenting the normalized positive and negative energy solution of the Klein-Gordon (KG) equation, we construct the thermal Hadamard function. Due to the  quasiperiodic condition obeyed by the quantum field along the compact dimension the corresponding moment  becomes discrete. So, in order to  develop the sum over this number to calculate the Hadamard function, we adopt the Abel-Plana summation formula. This formula allows to decompose this function as the sum of a part associated with the cosmic string without compactification, plus a contribution induced by the  compactification. In section \ref{Field_squared} we calculate the thermal expectation value of field squared, $\langle|\varphi|^2\rangle_T$, by taking directly the coincidence limit in the two-point thermal Hadamard function. Due to the decomposition of the thermal Hadamard function, the result obtained is also expressed as the sum of two contributions: one due to the cosmic sting without compactification plus the other induced by the comapctification.  Moreover, the thermal field squared in a even function of the chemical potential. Because $|\tilde{\mu}|\leq m$, being $m$ the mass associated to the quantum field, a non-vanishing value for the thermal field square can be provided in the limit of massless field.  Also in this section we present analytically the behaviors of  thermal expectation value of the field squared for some  asymptotic regimes of the parameters, such as low and high temperature. In addition we present some graphs, considering $D=3$, exhibiting the behavior of $\langle|\varphi|^2\rangle_T$ as function of temperature and distance to the string's core. In section \ref{Energy_momentum} we calculate the thermal expectation value of the energy-momentum tensor (EM), $\langle T^\mu_\nu\rangle_T$. This observable is also decomposed in terms of an uncompactied part plus the one induced by the compactification. Both contributions are even function of the chemical potential. This fact allows us to obtain non-vanishing values in the limit of massless field. Moreover, we show that the structure of the thermal energy-density, $\langle T^0_0\rangle_T$, is similar to the component along the compact dimension. This result is compatible with the Matsubara formalism where the thermal Hadamard function can be obtained from the zero-temperature one by imposing periodic function on the Euclidean time with period equal to $\beta=1/T$, being $T$ the temperature of the system. In this section we investigate in detail various asymptotic regime of the thermal energy density, including low and high temperature. Also we present some plots, considering $D=3$, exhibiting the behavior of  $\langle T^0_0\rangle_T$ as function of temperature and the distance to the string's core. Finally in section \ref{conc} we give a brief conclusions of our most relevant results. Throughout the paper we use natural units $G=\hbar =c=k_B=1$.

\section{Model setup and thermal Hadamard function}
\label{sec2}
The main objective of this section is to obtain the thermal Hadamard function associated to a charged bosonic quantum field 
propagating in the $(1+D)-$dimensional compactified cosmic string spacetime, with $D\geq 3$. In order to do that we present first line element associated with the space background under consideration, having a cosmic string along $z$-axis. Using cylindrical coordinates, the line element is expressed by 
\begin{equation}
	ds^{2}=g_{\mu\nu}dx^{\mu}dx^{\nu}=dt^{2}-dr^2-r^2d\phi^2-dz^2- \sum_{i=4}^{D}(dx^{i})^2 \ .
	\label{ds1}
\end{equation}
where we assume for this coordinate system the following ranges: $r\geq 0$, $0\leq\phi\leq \phi_0= 2\pi/q$ and $-\infty< (t, \ x^i) < +\infty$ for $i=4,...,D$.  The coordinate $z$ is compactified to a circle with length $L$, so $z\in[0, \ L ]$.
The presence of the cosmic string is codified through the parameter $q\geq 1$.

In the presence of a gauge field, $A_{\mu }$, the field equation that  governs the quantum dynamics of the charged massive bosonic field in a curved background is
\begin{equation}
	\left({\mathcal{D}}^2+m^2+\xi R\right)\varphi(x)=0  \  ,  \label{K-G}
\end{equation}
where the differential operator in the field equation reads
\begin{align}
	{\mathcal{D}}^2=\frac{1}{\sqrt{|g|}}{\mathcal{D}}_{\mu}\left(\sqrt{|g|}g^{\mu \nu} {\mathcal{D}}_{\nu
	}\right), \ {\mathcal{D}}_{\mu
	}=\partial _{\mu }+ieA_{\mu }\   \ {\rm with} \  \  g=\det(g_{\mu\nu})  \  .  \label{1}
\end{align}
Moreover, we have considered the presence of a non-minimal coupling, $\xi$, between the field and the geometry represented by the Ricci scalar, $R$.

An important step to calculate the Hadamard function is to present the complete set of normalized positive/negative energy solutions  of \eqref{K-G}. In our analysis we will admit that the scalar field obeys the quasiperiodic condition along the $z$-axis,
\begin{equation}
	\varphi (t,r,\phi,z+L,x^4,...,x^D)=e^{2\pi i\eta}\varphi(t,r,\phi,z,x^4,...,x^D) \ ,  
	\label{QPC}
\end{equation}
with a constant phase $\eta\in[0, \ 1] $. In addition, we consider the interaction of the charged field with two magnetic fields, one along the string's core and the other enclosed by the compactified coordinate. These magnetic fields are represented  by the two independent components of the vector potentials,
\begin{equation}
	A_{\mu}=(0,0,{-q\Phi_\phi}/{2\pi},{-\Phi_z}/{L}, 0 \ , \ ... \ 0)\ . 
	\label{Vector}
\end{equation}
In the above expression, $\Phi_\phi$ and $\Phi_z$ correspond to the magnetic fluxes along the string's core and enclosed by the compactified direction, respectively.  

In quantum field  theory the  condition \eqref{QPC} changes the spectrum of the vacuum fluctuations compared with the case of uncompactified dimension, consequently inducing new contributions to the average of relevant physical observables.

In the geometry defined by \eqref{ds1} and in the presence of the vector  potentials given above, the KG equation \eqref{K-G} becomes
\begin{equation}
	\left[\partial_t^2-\partial_r^2-\frac{1}{r}\partial_r-\frac{1}{r^2}(\partial_{\phi}+
	ieA_{\phi})^2-(\partial_{z}+ieA_{z})^2-\sum_{i=4}^{D}\partial_{i}^{2}+m^2\right]\varphi(x)=0 \ . 
	\label{KG_1}
\end{equation}

The normalized positive/negative energy wave function  solution of \eqref{KG_1} was derived in \cite{Braganca_15}. It reads,
\begin{equation}
	\varphi_{\sigma}^{(\pm)}(x)=\left(\frac{q\lambda}{2(2\pi)^{D-2}E_\sigma L}\right)^{\frac{1}{2}} 
	J_{q|n+\alpha|}(\lambda r)e^{\mp iE_\sigma t+iqn\phi+ik_l z+i{\vec{k}\cdot{\vec{x}}}} \ .
	\label{Solu}
\end{equation}
Where  $J_\mu(z)$ corresponds to the Bessel function \cite{Abra} and  $\vec{x}$ the coordinates defined in the $(D-4)$ extra dimensions, being $\vec{k}$ the corresponding momentum. Moreover, in the expression above,  $\sigma$ represents the set of quantum numbers $(n, \lambda, k_l, \bold k_{||})$, being $n=0,\pm1,\pm2,\ldots$, $\lambda \geq 0$, $-\infty<k^j<\infty$ for $j=4,...,D$. The order of Bessel function is characterized by the presence of the parameter $\alpha$, defined by
\begin{eqnarray}
\alpha =eA_{\phi }/q=-\Phi_{\phi }/\Phi _{0}  \  .   \label{betaj}
\end{eqnarray}
being $\Phi _{0}=2\pi /e$ the quantum flux.

 The momentum along the string axis, $k_l$, is discrete due to the compactification condition, Eq. \eqref{QPC}. It reads, 
\begin{eqnarray}
	k_l=2\pi(l+\eta)/L \  , \ {\rm with} \ l=0,\pm 1,\pm 2\ ,\ldots  
\end{eqnarray}
The energy is expressed in terms of $\lambda $, $\vec{k}$ and $l$ by the relation
\begin{equation}
	E_\sigma=\sqrt{\lambda ^{2}+\tilde{k}_{l}^{2}+\vec{k}^2+m^{2}}\ ,\;\tilde{k}_{l}=2\pi (l+
	\tilde{\eta})/L \  ,  \label{E+}
\end{equation}
where
\begin{equation}
	\tilde{\eta}=\eta +eA_{z}L/(2\pi )=\eta -\Phi _{z}/\Phi _{0}\ .
	\label{bett}
\end{equation}

Assuming that the field is in equilibrium state with temperature $T=1/\beta$, the thermal Hadamard function is defined as:
\begin{eqnarray}
	\label{Hada}
	G(x,x')=\text{Tr}\left[\hat{\rho}\left(\hat{\varphi}(x)\hat{\varphi}^\dagger(x')+\hat{\varphi}^\dagger(x')\hat{\varphi}(x)\right)\right] \  , 
\end{eqnarray}
where $\hat{\rho}$ is density matrix defined as,
\begin{eqnarray}
	\label{rho}
\hat{\rho}=Z^{-1}e^{-\beta(\hat{H}-\mu'\hat{Q})} \  .	
\end{eqnarray}
In the above equation $\hat{H}$ is the Hamiltonian operator, $\hat{Q}$ the conserved charge, and $\mu ^{\prime }$ the corresponding chemical potential. The grand canonical partition function, $Z$, is,
\begin{equation}
	Z=\mathrm{tr}[e^{-\beta (\hat{H}-\mu ^{\prime }\hat{Q})}] \ . \label{partition}
\end{equation}

Expanding the field operator in terms of the complete set of normalized positive/negative energy solutions of \eqref{K-G},
\begin{eqnarray}
	\label{phi}
\hat{\varphi}(x)=\sum_\sigma[\hat{a}_\sigma\varphi_\sigma^{(+)}(x)+\hat{b}_\sigma^\dagger\varphi^{(-)}_\sigma(x)] \  ,  
\end{eqnarray}
where the summation over $\sigma$ represents,
\begin{eqnarray}
	\label{Sum}
	\sum_\sigma=\int_0^\infty d\lambda\int d{\vec{k}}\sum_n\sum_l  \  ,
\end{eqnarray}
and using the following relations
\begin{eqnarray}
	\mathrm{tr}[\hat{\rho }\hat{a}_{\sigma }^{+}\hat{a}_{\sigma ^{\prime }}] &=&%
	\frac{\delta _{\sigma \sigma ^{\prime }}}{e^{\beta (E_l-\tilde{\mu} )}-1},  \notag \\
	\mathrm{tr}[\hat{\rho }\hat{b}_{\sigma }^{+}\hat{b}_{\sigma ^{\prime }}] &=&%
	\frac{\delta _{\sigma \sigma ^{\prime }}}{e^{\beta (E_l+\tilde{\mu} )}-1} ,  \label{traa}
\end{eqnarray}%
being $\tilde{\mu}=e \mu'$. After many intermediate steps, we can express the Hadamard function as,
\begin{eqnarray}
	\label{Hada1}
	G(x,x')=G_0(x,x')+G_T(x,x') \ .
\end{eqnarray}
In the equation above, the first term corresponds the Hadamard function at zero temperature and the second contribution, $G_T(x,x')$, the thermal correction.

\subsection{Thermal Hadamard function}
The Hadamard function at zero temperature has been obtained in \cite{Braganca_19}. 
So  our main objective in this subsection is to develop the calculation of the thermal Hadamard function.

In a compact notation the thermal Hadamard function reads,
\begin{eqnarray}
	\label{Thermal_1}
	G_T(x,x')=2\sum_{s=\pm}\sum_{\sigma}\varphi^{(s)}_\sigma(x)(\varphi^{(s)}_\sigma(x')) ^*\frac1{e^{\beta(E_\sigma-s\tilde{\mu})}-1}  \  .
\end{eqnarray}
Substituting the expressions for the normalized solution of the Klein-Gordon equation, Eq. \eqref{Solu}, we have,
\begin{eqnarray}
	\label{Thermal_2}
	G_T(x,x')&=&\frac{q}{(2\pi)^{D-2}L}\sum_\sigma\lambda J_{q|n+\alpha|}(\lambda r)J_{q|n+\alpha|}(\lambda r')\frac{e^{iqn\Delta\phi+ik_l\Delta z+i\vec{k}\cdot\Delta{\vec{x}}}}{E_\sigma}\nonumber\\
	&\times&\left[\frac{e^{-iE_\sigma\Delta t}}{e^{\beta(E_\sigma-\tilde{\mu} )}-1}+\frac{e^{iE_\sigma\Delta t}}{e^{\beta(E_\sigma+\tilde{\mu} )}-1}\right] \  .
\end{eqnarray}

Notice that the bosonic chemical potential is restricted to the condition $|\tilde{\mu}|\leq E_0$, being $E_0$ the minimal energy of the system. Therefore $E_\sigma\pm \tilde{\mu}$ is always a positive quantity. This enables us to adopt the series expansion below, 
\begin{eqnarray}
	\label{Ident}
	\frac1{e^y-1}=\sum_{j=1}e^{-jy}  \  . 
\end{eqnarray}
in the development of \eqref{Thermal_2}. Doing a Wick rotation, $t=i\tau$, we can express the thermal Hadamard function as,
\begin{eqnarray}
	\label{Thermal_3}
	G_T(x,x')&=&\frac{q\delta(\Delta z)}{(2\pi)^{D-2}L}\sum_{j=1}^\infty\int d{\vec{k}}e^{i\vec{k}\cdot\Delta{\vec{x}}}\sum_{n=-\infty}^\infty e^{iqn\Delta\phi}\int_0^\infty  d\lambda\lambda   J_{q|n+\alpha|}(\lambda r)J_{q|n+\alpha|}(\lambda r')\nonumber\\
	&\times&\sum_{l=-\infty}^\infty\frac{e^{i{\tilde{k}}_l\Delta z}}{E_\sigma}
	\left[e^{j\beta\tilde\mu}e^{-E_\sigma(j\beta-\Delta\tau)}+e^{-j\beta{\tilde\mu}}
	e^{-E_\sigma(j\beta+\Delta\tau)}\right] \  ,
\end{eqnarray}
where we have introduced a function $\delta(\Delta z)=e^{- ie A_z \Delta z}$, in order to write the exponential dependence of $k_l$ as
\begin{equation}
	e^{ik_z \Delta z}=e^{2\pi i(l+\eta)\Delta z/L}=e^{i\tilde{k}_z \Delta z}e^{2\pi i\Phi_z \Delta z/(L\Phi_0)}=
	e^{i\tilde{k}_z \Delta z}\delta(\Delta z)  \  . 
\end{equation}

For the evaluation of the sum over the quantum number $l$ in \eqref{Thermal_3} we will use the Abel-Plana summation formula given in the form \cite{Saha_2010,SahaRev}
\begin{eqnarray}
	\sum_{l=-\infty }^{\infty }g(l+\tilde{\eta} )f(|l+\tilde{\eta} |)&=&\int_{0}^{\infty }du
	\left[ g(u)+g(-u)\right] f(u) + i\int_{0}^{\infty }du\left[ f(iu)-f(-iu)\right]\nonumber\\
	&\times&\left(\frac{g(i u)}{e^{2\pi (u+i \tilde{\eta} )}-1}+\frac{g(-i u)}{e^{2\pi (u-i \tilde{\eta} )}-1}\right)  \  .
	\label{sumform}
\end{eqnarray}
For our case, we have
\begin{equation}
	g(u)= e^{(2\pi i u/{L}) \Delta z} \ \ \ {\rm and} \ \ \ f(u)=\frac{e^{-\sigma_j^{(\pm)}\sqrt{p^2+({2\pi u}/{L})^2}}}
	{\sqrt{p^2+({2\pi u}/{L})^2}} \  ,
\end{equation}
with $p=\sqrt{m^2+\lambda^2+{\vec{k}}^2}$, $\sigma_j^{(\pm)}=j\beta\pm\Delta\tau$ and  $u=l+\tilde\eta$. Doing this, we can write the Hadamard function in the decomposed form
\begin{equation}
	G_T(x,x')=G_{Ts}(x,x')+G_c(x,x')  \  ,
	\label{decomposed}
\end{equation}
where $G_{Ts}(x,x')$ corresponds to the contribution in the cosmic string background without the compactification coming from the first term on the right hand side of Eq. \eqref{sumform}, and $G_{c}(x,x')$ is the contribution due the compactification and comes from the second term on the right hand side.

For the contribution due the cosmic string and after some intermediate steps, we get,
\begin{eqnarray}
	G_{Ts}(x,x')&=&\frac{2q\delta(\Delta z)}{(2\pi)^{D-1}}\sum_{j=1}^\infty \int d\vec{k} \ e^{i\vec{k}\cdot \Delta\vec{x}}\sum_{n=-\infty}^\infty e^{inq\Delta\phi}
	\int_0^\infty d\lambda \ \lambda J_{q|n+\alpha|}(\lambda r)J_{q|n+\alpha|}(\lambda r')\nonumber\\
	&\times&\left(e^{j\beta\tilde{\mu}}\int_0^\infty dv \cos(v\Delta z)\frac{e^{-\sigma_j^{(-)}\sqrt{p^2+(2\pi u/L)^2}}}{\sqrt{p^2+(2\pi u/L)^2 }}\right.\nonumber\\
	&+&\left.e^{-j\beta\tilde{\mu}}\int_0^\infty dv \cos(v\Delta z)\frac{e^{-\sigma_j^{(+)}\sqrt{p^2+(2\pi u/L)^2}}}{\sqrt{p^2+(2\pi u/L)^2 }}\right) \  ,
	\label{Thermal_s}
\end{eqnarray}
where we have defined a new variable $v=2\pi u/L$.

Using the identity,
\begin{eqnarray}
	\label{identity2}
	\frac{e^{-\sigma\sqrt{p^2+v^2}}}{\sqrt{p^2+v^2}}=\frac2{\sqrt{\pi}}\int_0^\infty ds e^{-(p^2+v^2)s^2-\sigma^2/(4s^2) }  \  ,
\end{eqnarray}
It is possible to integrate over $\lambda$ and $v$ in \eqref{Thermal_s}, by using \cite{Grad}. Defining a new variable $u=1/(2s^2)$, and after some intermediate steps, we get,
\begin{eqnarray}
	\label{Thermal_s_1}
	G_{Ts}(x,x')&=&\frac{q\delta(\Delta z)}{(2\pi)^{\frac{D+1}2}}\sideset{}{'}\sum_{j=-\infty}^\infty e^{j\beta\tilde{\mu}}\int_0^\infty du u^{\frac{D-3}{2}}e^{-{m^2}/{(2u)}}e^{-({\cal{V}}^2+(\Delta\tau-j\beta)^2)u/2} {\cal{I}}(q,\alpha,\Delta\phi, urr') 
	 \   , \nonumber\\
\end{eqnarray}
where the prime in the summation over $j$ means that the term $j=0$ should be excluded, 
\begin{eqnarray}
	\label{V}
	{\cal{V}}^2=r^2+r'^2+\Delta{\vec{x}} ^2+\Delta z^2
\end{eqnarray}
and 
\begin{equation}
	\mathcal{I}(q,\alpha,\Delta\phi,urr')=\sum_{n=-\infty}^\infty e^{inq\Delta\phi}I_{q|n+\alpha|}(urr') \  ,
	\label{nsum}
\end{equation}
being $I_\nu(z)$ the modified Bessel function \cite{Grad}.

We can obtain a more convenient  expression for \eqref{Thermal_s_1} writing the parameter $\alpha$ in \eqref{betaj} in the form
\begin{equation}
	\alpha=n_0 +\alpha_0 \ \ {\rm with  \ |\alpha_0| \ < 1/2}  \  ,
	\label{alphazero}
\end{equation}
being $n_0$ an integer number. Moreover, we will use the expression below for \eqref{nsum}, obtained in \cite{deMello:2014ksa}:
\begin{eqnarray}
	\mathcal{I}(q, \alpha,\Delta\phi,x)&=&\frac{1}{q}\sum_k e^{x\cos(2k\pi /q-\Delta\phi)}e^{i\alpha(2k\pi -q\Delta\phi)}-
	\frac{e^{-iqn_0\Delta\phi}}{2\pi i}\sum_{\chi=\pm 1}\chi e^{i\chi \pi q\alpha_0}\nonumber\\
	&\times&\int_0^\infty dy \frac{\cosh[qy(1-\alpha_0)]-\cosh(q\alpha_0y)e^{-iq(\Delta\phi+\chi\pi)})}{e^{x\cosh y}[\cosh(qy)-\cos(q(\Delta\phi+\chi\pi))]}.
	\label{Sum_n}
\end{eqnarray}
For the summation over $k$ we have the condition
\begin{equation}
	-\frac{q}{2}+\frac{2\pi}{q}\Delta\phi \leq k \leq \frac{q}{2}+\frac{2\pi}{q}\Delta\phi  \  .
	\label{conditionk}
\end{equation}

Substituting \eqref{Sum_n} into \eqref{Thermal_s_1}, we obtain,
\begin{eqnarray}
	\label{Thermal_s_2}
G_{Ts}(x,x')&=&\frac{\delta(\Delta z)}{(2\pi)^{\frac{D+1}{2}}}\sideset{}{'}\sum_{j=-\infty}^\infty e^{j\beta\tilde{\mu}}\int_0^\infty du u^{\frac{D-3}{2}}e^{-{m^2}/{(2u)}}e^{-({\cal{V}}^2+(\Delta\tau-j\beta)^2)u/2}\nonumber\\
&\times& \left\{\sum_k e^{urr'\cos(2k\pi /q-\Delta\phi)}e^{i\alpha(2k\pi -q\Delta\phi)}-
\frac{qe^{-iqn_0\Delta\phi}}{2\pi i}\sum_{\chi=\pm 1}\chi e^{\chi i\pi q\alpha_0}\right.\nonumber\\
&\times&\left.\int_0^\infty dy \frac{e^{-urr'\cosh y}[\cosh[qy(1-\alpha_0)]-\cosh(q\alpha_0y)e^{-iq(\Delta\phi+\chi\pi)})]}{\cosh(qy)-\cos(q(\Delta\phi+\chi\pi))}\right\} \ . 
\end{eqnarray}

For further manipulation of the above expression, we employ the integral representation for the Macdonald function \cite{Grad}, 
\begin{eqnarray}
\label{Macd}
K_\nu(xz)=\frac{z^\nu}{2}\int\frac{dt}{t^{\nu+1}}e^{-xt/2-xz^2/(2t)}  \  . 
\end{eqnarray}
It allows us to express \eqref{Thermal_s_2} as, 
\begin{eqnarray}
\label{Thermal_s_3}
G_{Ts}(x,x')&=&\frac{2m^{D-1}\delta(\Delta z)}{(2\pi)^{\frac{D+1}{2}}}\sideset{}{'}\sum_{j=-\infty}^\infty e^{j\beta\tilde{\mu}}\left\{\sum_k e^{i\alpha(2\pi k-\Delta\phi)} f_{\frac{D-1}{2}}\left(m{\bar{\cal{V}}}_k\right)\right.\nonumber\\
&-&\frac{qe^{-iqn_0\Delta\phi}}{2\pi i}\sum_{\chi=\pm 1}\chi e^{\chi i\pi q\alpha_0}\int_0^\infty dy f_{\frac{D-1}{2}}\left(m{\bar{\cal{V}}}_y\right)\nonumber\\
&\times&\left.\frac{[\cosh[qy(1-\alpha_0)]-\cosh(q\alpha_0y)e^{-iq(\Delta\phi+\chi\pi)})]}{\cosh(qy)-\cos(q(\Delta\phi+\chi\pi))}\right\}  \  ,
\end{eqnarray}
where we are using the notation,
\begin{eqnarray}
	\label{V_ky}
	{\bar{\cal{V}}}_k&=&\sqrt{ r^2+r'^2-2rr'\cos(2k\pi/q-\Delta\phi)+\Delta{\vec{x}} ^2+\Delta z^2 +(\Delta\tau-j\beta)^2} \ ,  \nonumber\\
		{\bar{\cal{V}}}_y&=& \sqrt{r^2+r'^2+2rr'\cosh(y)+\Delta{\vec{x}} ^2+\Delta z^2 +(\Delta\tau-j\beta)^2}	\ .
\end{eqnarray}
Moreover, in \eqref{Thermal_s_3} we have defined,
\begin{eqnarray}
	\label{f-function}
	f_\nu(z)=\frac{K_\nu(z)}{z^\nu} \  .
\end{eqnarray}

The contribution induced by the compactification on the thermal Hadamard function is expressed by,
\begin{eqnarray}
	\label{GTc}
	G_{Tc}&=&\frac{q i\delta(\Delta z)}{(2\pi)^{D-1}}\sum_{j=1}^\infty \int d\vec{k}  e^{i\vec{k}\cdot \Delta\vec{x}}\sum_{n=-\infty}^\infty e^{inq\Delta\phi}
		\int_0^\infty d\lambda \ \lambda J_{q|n+\alpha|}(\lambda r)J_{q|n+\alpha|}(\lambda r')\nonumber\\
		&\times&\sum_{\delta=\pm1}e^{-j\delta\beta\tilde{\mu}}\int_0^\infty dv\left[\frac{e^{-\sigma_j^{(\delta)}\sqrt{p^2+(iv)^2}}}{\sqrt{p^2+(iv)^2}}-\frac{e^{-\sigma_j^{(\delta)}\sqrt{p^2+(-iv)^2}}}{\sqrt{p^2+(-iv)^2}}\right]\nonumber\\
		&\times&\left(\frac{e^{-v\Delta z}}{e^{Lv+2\pi i \tilde{\eta}}-1}+\frac{e^{v\Delta z}}{e^{Lv-2\pi i \tilde{\eta}}-1}\right)  \  ,
\end{eqnarray}
where we have made a change of variable, $v=2\pi u/L$. In the above equation $p=\sqrt{m^2+\lambda^2+{\vec{k}}^2}$ and $\sigma^{(\pm)}= j\beta\pm \Delta\tau$.

The integral over the $u$ variable must be considered in two intervals: for $v<p$ and $v>p$. In the first interval the integral vanishes;  however for the second interval we use the identity $\sqrt{(\pm iv)^2+p^2}=\pm i\sqrt{v^2-p^2}$. So, we
can write,
\begin{eqnarray}
	\label{GTc1}
		G_{Tc}&=&\frac{4q \delta(\Delta z)}{(2\pi)^{D-1}}\sideset{}{'}\sum_{j=-\infty}^\infty e^{j\beta\tilde{\mu}} \int d\vec{k}  e^{i\vec{k}\cdot \Delta\vec{x}}\sum_{n=-\infty}^\infty e^{inq\Delta\phi}
	\int_0^\infty d\lambda \ \lambda J_{q|n+\alpha|}(\lambda r)J_{q|n+\alpha|}(\lambda r')\nonumber\\
	&\times&\int_p^\infty dv \frac{\cos[(j\beta-\Delta \tau)\sqrt{v^2-p^2}]}{\sqrt{v^2-p^2}}\sum_{l=1}^\infty e^{-lLv}
	\cosh(v\Delta z +2\pi i l\tilde{\eta})  \  , 
\end{eqnarray}
where we have used the identity \eqref{Ident} to express the last term in \eqref{GTc}.

Briefly in this paragraph, we present a few steps taken to obtain a more workable expression for \eqref{GTc1}. First we introduce a new variable $w^2=v^2-p^2$, using the identity \eqref{identity2} it is possible to integrate over the variable $\lambda$ and the integral over the new variable $w$. Finally by a convenient changing of variable, we find,
\begin{eqnarray}
	\label{GTc2}
	G_{Tc}&=&\frac{q \delta(\Delta z)}{(2\pi)^{\frac{D+1}2}}\sideset{}{'}\sum_{j=-\infty}^\infty e^{j\beta\tilde{\mu}} \sideset{}{'}\sum_{l=-\infty}^\infty e^{2\pi il\tilde{\eta}} \int_0^\infty du u^{\frac{D-3}{2}}e^{-m^2/(2u)} {\cal{I}}(q,\Delta\phi,urr') e^{-({{\cal{V}}^{(l)}_j})^2u/2} \  ,  \nonumber\\
\end{eqnarray}
being
\begin{eqnarray}
	\label{V_l}
	\left({{\cal{V}}^{(l)}_j}\right)^2=r^2+r'^2+\Delta{\vec{x}}^2+(\Delta\tau-j\beta)^2+(\Delta z-lL)^2  \ .
\end{eqnarray}
Substituting \eqref{Sum_n} into \eqref{GTc2} and using the integral representation for the Macdonald function, Eq. \eqref{Macd}, we can express the thermal Hadamard function induced by the compactification in the form:
\begin{eqnarray}
	\label{GTc3}
	G_{Tc}(x,x')&=&\frac{2m^{D-1}\delta(\Delta z)}{(2\pi)^{\frac{D+1}{2}}}\sideset{}{'}\sum_{j=-\infty}^\infty e^{j\beta\tilde{\mu}}\sideset{}{'}\sum_{l=-\infty}^\infty e^{2\pi il\tilde{\eta}}\left\{\sum_k e^{i\alpha(2\pi k-\Delta\phi)} f_{\frac{D-1}{2}}\left(m{{\cal{V}}^{(l)}_{jk}}\right)\right.\nonumber\\
	&-&\frac{qe^{-iqn_0\Delta\phi}}{2\pi i}\sum_{\chi=\pm 1}\chi e^{\chi i\pi q\alpha_0}\int_0^\infty dy f_{\frac{D-1}{2}}\left(m{{\cal{V}}^{(l)}_{jy}}\right)\nonumber\\
	&\times&\left.\frac{[\cosh[qy(1-\alpha_0)]-\cosh(q\alpha_0y)e^{-iq(\Delta\phi+\chi\pi)})]}{\cosh(qy)-\cos(q(\Delta\phi+\chi\pi))}\right\}  \  ,
\end{eqnarray}
with
\begin{eqnarray}
	\label{V_2}
{{\cal{V}}^{(l)}_{jk}}&=&\sqrt{r^2+r'^2-2rr'\cos(2\pi k/q-\Delta\phi)+\Delta{\vec{x}}^2+(\Delta\tau-j\beta)^2+(\Delta z-lL)^2}  \nonumber\\
{{\cal{V}}^{(l)}_{jy}}&=&\sqrt{r^2+r'^2+2rr'\cosh(y)+\Delta{\vec{x}}^2+(\Delta\tau-j\beta)^2+(\Delta z-lL)^2}  \ 
\end{eqnarray}
and adopting the notation \eqref{f-function}.

A compact notation for the total thermal Hadamard function, Eq. \eqref{decomposed}, can be provided by taking our previous result for the contribution due to the uncompactified cosmic string spacetime part, Eq. \eqref{Thermal_s_3}, plus the compactified one, Eq. \eqref{GTc3}. This expression is,
\begin{eqnarray}
	\label{Hadamard_thermal}
	G_{T}(x,x')&=&\frac{2m^{D-1}\delta(\Delta z)}{(2\pi)^{\frac{D+1}{2}}}\sideset{}{'}\sum_{j=-\infty}^\infty e^{j\beta\tilde{\mu}}\sum_{l=-\infty}^\infty e^{2\pi il\tilde{\eta}}\left\{\sum_k e^{i\alpha(2\pi k-\Delta\phi)} f_{\frac{D-1}{2}}\left(m{\cal{V}}^{(l)}_{jk}\right)\right.\nonumber\\
	&-&\frac{qe^{-iqn_0\Delta\phi}}{2\pi i}\sum_{\chi=\pm 1}\chi e^{\chi i\pi q\alpha_0}\int_0^\infty dy f_{\frac{D-1}{2}}\left(m{{\cal{V}}^{(l)}_{jy}}\right)\nonumber\\
	&\times&\left.\frac{[\cosh[qy(1-\alpha_0)]-\cosh(q\alpha_0y)e^{-iq(\Delta\phi+\chi\pi)})]}{\cosh(qy)-\cos(q(\Delta\phi+\chi\pi))}\right\}  \  ,
\end{eqnarray}
with the same ${{\cal{V}}^{(l)}_{jk}}$ and ${{\cal{V}}^{(l)}_{jy}}$ given in \eqref{V_2}, but now considering $l=0$. In fact the component $l=0$ corresponds to the uncompactified thermal Hadamard function.

This is the final and most compact expression for the total Hadamard function and it allows us to present the VEVs of the field squared and the energy-momentum tensor for a charged scalar massive quantum field in a closed form for a general value of $q$.

\section{Thermal expectation value of the field squared}
\label{Field_squared}
In this section we calculate the thermal correction to the expectation value of the field squared. This quantity is relevant to evaluation of the VEV of the thermal energy-momentum tensor. The expectation value of the field squared is formally given by
\begin{eqnarray}
	\langle|\varphi|^2\rangle=\frac12\lim_{x'\to x}G(x,x') \  . 
\end{eqnarray}
Due to the decomposition \eqref{Hada1}, the expectation value of the field squared is expressed as the sum of the zero-temperature contribution plus the contribution associated with particle  and anti-particle, as shown below,
\begin{eqnarray}
		\langle|\varphi|^2\rangle=\langle|\varphi|^2\rangle_0+	\langle|\varphi|^2\rangle_T  \  ,
\end{eqnarray}
where the first term on the right hand side of the above equation, represents the zero-temperature vacuum expectation value (VEV) contribution to the field squared, and the second term is the thermal correction. The zero-temperature contribution has already been obtained and analyzed in \cite{Braganca_19}. So our focus here is in the obtainment of the thermal correction. The thermal correction, on the other hand, can be decomposed as the sum of the part due to the uncompactified cosmic string spacetime, plus the contribution induced by the compactification. These are obtained by taking the coincidence limit of the \eqref{Thermal_s_3} and \eqref{GTc3}, respectively\footnote{In this analysis we can take coincidence limit in the thermal Hadamard function directly. No divergence will take place. The divergence is presented in the zero-temperature term, where a normalization procedure is required.}. In this sense we can write,
\begin{eqnarray}
	\label{phi2}
\langle|\varphi|^2\rangle_T  =\langle|\varphi|^2\rangle_{Ts}+ \langle|\varphi|^2\rangle_{Tc}  \  .
\end{eqnarray}

Let us first analyze the uncompactified term. It is given by the expression below,
\begin{eqnarray}
	\label{phi2s}
\langle|\varphi|^2\rangle_{Ts}&=&\frac{4m^{D-1}}{(2\pi)^{\frac{D+1}{2}}}\sum_{j=1}^\infty\cosh(j\beta\tilde{\mu})\left\{\sum_{k=1}^{[\frac q2]}\cos(2k\pi\alpha_0) f_{\frac{D-1}{2}}\left(mu_{jk}\right)\right.\nonumber\\
&-&\left.\frac{q}{2\pi}\int_0^\infty dy f_{\frac{D-1}{2}}\left(mu_{jy}\right)\frac{h(q,\alpha_0,y)}{\cosh(qy)-\cos(q\pi)}\right\}  \  ,
\end{eqnarray}
where we use the notations,
\begin{eqnarray}
	\label{h}
	h(q,\alpha_0,y)&=&\cosh[qy(1-\alpha_0)]\sin(q\pi\alpha_0)+\cosh(q\alpha_o y)\sin{q\pi(1-\alpha_0)} \ , \nonumber\\
u_{jk}&=&\sqrt{4r^2\sin^2(k\pi/q)+(j\beta)^2} \ , \nonumber\\
	u_{jy}&=&\sqrt{4r^2\cosh(y/2)+(j\beta)^2} \ .
\end{eqnarray}
In the above expression $[\frac q2]$ means the biggest integer part of $q/2$. Moreover, note that we have omitted the term $k=0$ in \eqref{phi2s}, which is the thermal contribution in Minkowski spacetime, i.e., in the absence of the string, and it is explicitly given by
\begin{equation}
	\label{phi2_M_s}
	\langle|\varphi|^2\rangle_{T}^{(M)}=\frac{2m^{D-1}}{(2\pi)^{\frac{D+1}{2}}}\sum_{j=1}^\infty\cosh(j\beta\tilde{\mu})f_{\frac{D-1}{2}}(mu_{j0}),
\end{equation}
where $u_{j0}=j\beta$. This contribution is always positive. Since our goal in this paper is the study of the contributions induced by the string and by the compactification, we will discard the analysis of this term.
 
 For the massless field case, the field squared is simplified and takes the form,
\begin{eqnarray}
	\label{phi2s0}
	\langle |\varphi|^2\rangle_{Ts}&=&\frac{\Gamma\left(\frac{D-1}{2}\right)}{\pi^{\frac{D+1}{2}}}\sum_{j=1}^{\infty}\Bigg\{\sum_{k=1}^{[\frac q2]}\frac{\cos(2k\pi\alpha_0)}{[4r^2\sin^2(\pi k/q)+(j\beta)^2]^{(D-1)/2}} \nonumber \\
	&-&\frac{q}{2\pi}\int_0^\infty \frac{dy}{[4r^2\cosh^2(y/2)+(j\beta)^2]^{(D-1)/2}}\frac{h(q,\alpha_0,y)}{\cosh(qy)-\cos(q\pi)}\Bigg\}  \  .
\end{eqnarray}

Let us analyze the limit $mr\gg1$ in \eqref{phi2s}. For $q>2$, the dominant contribution comes from the term with $k=1$ and $j=1$:
\begin{equation}
	\langle |\varphi|^2\rangle_{Ts}
	\approx\frac{m^{\frac{D}{2}-1}}{(2\pi)^{D/2}}e^{\beta|\tilde{\mu}|}\frac{e^{-m\sqrt{4r^2\sin^2(\pi /q)+\beta^2}}}{(4r^2\sin^2(\pi k/q)+\beta^2)^{D/4}} \ .
\end{equation}
On the other hand, for $1\le q<2$, the dominant contribution is:
\begin{equation}
	\langle |\varphi|^2\rangle_{Ts}
	\approx-\frac{qm^{\frac{D}{2}-1}}{(2\pi)^{D/2+1}}\frac{e^{\beta|\tilde{\mu}|-m\sqrt{4r^2+\beta^2}}}{(4r^2+\beta^2)^{D/4}}\int_{0}^{\infty}dy\frac{h(q,\alpha_0,y)}{(\cosh(qy)-\cos(q\pi))} \ .
\end{equation}

In the low temperature regime, $T\ll m,r^{-1}$, the parameter $\beta$ is large and since it is present in the argument of the function $f_{\nu}(x)$, we can use the asymptotic expansion of the Macdonald function for large arguments, $K_{\nu}(x)\approx\left(\frac{\pi}{2x}\right)^{1/2}e^{-x}$. In this case, the dominant contribution comes from the term $j=1$ and we can set $u_{1k}\approx \beta$ in the summation over $k$ and $u_{1y}\approx \beta$ in the integral over $y$. After some intermediate steps, we obtain,
\begin{eqnarray}
	\label{T_low}
	\langle|\varphi|^2\rangle_{Ts}&\approx&\frac{m^{\frac D2-1}g(q,\alpha_0)}{(2\pi)^{\frac{D}{2}}}T^{D/2}e^{-(m-|\tilde{\mu}|)/T}  \  ,
\end{eqnarray}
with $g(q,\alpha_0)$ being given by
\begin{equation}
	\label{g-function}
	g(q,\alpha_0)=\sum_{k=1}^{[\frac q2]}\cos(2k\pi\alpha_0)-\frac{q}{2\pi}\int_0^\infty dy\frac{h(q,\alpha_0,y)}{\cosh(qy)-\cos(q\pi)} \ .
\end{equation}
We clearly see that \eqref{T_low} goes to zero, since $|\tilde{\mu}|\le m$.

On the other hand, at the high temperature limit, $T\gg m,r^{-1}$, the main contribution to the field squared comes from large $j$, and consequently representation \eqref{phi2s} is not convenient for this limiting case. In order to evaluate this behaviour, we write $\cosh(j\beta \tilde{\mu})=\cos(ij\beta\tilde{\mu})$ and use the identity below \cite{Bellucci:2009jr, Cruz:2018bqt}
\begin{eqnarray}
	\label{Resum}
	&&\sum_{j=1}^{\infty}\cos(\alpha j )f_{\nu}\left(m\sqrt{b^2+\beta^2j^2}\right)=-\frac{1}{2}f_{\nu}\left(mb\right)+\frac{1}{2}\frac{\sqrt{2\pi}}{\beta m^{2\nu}}\sum_{j=-\infty}^{\infty}w_{j}^{2\nu-1}f_{\nu-1/2}(bw_{j}) \ ,
\end{eqnarray}
where $b>0$ and $w_{j}=\sqrt{(2\pi j+\alpha)^2/\beta^2+m^2}$. For our case, we have $\alpha=i\beta\tilde{\mu}$,  $b=2r{\small \begin{cases}
		\sin(\pi k/q) \\
		\cosh(y/2)
\end{cases}}$, $\nu=\frac{D-1}{2}$ and $w_{j}=\sqrt{(2\pi j/\beta+i\tilde{\mu})^2+m^2}$. Therefore, the leading term is $j=0$, so we obtain
\begin{eqnarray}
	\label{T_high}
	\langle|\varphi|^2\rangle_{Ts}&\approx&\frac{2(m^2-\tilde{\mu}^2)^{\frac{D}{2}-1}T}{(2\pi)^{\frac{D+1}{2}}}\left\{\sum_{k=1}^{[\frac q2]}\cos(2k\pi\alpha_0) f_{\frac{D}{2}-1}\left(2r\sin(\pi k/q)\sqrt{m^2-\tilde{\mu}^2}\right)\right.\nonumber\\
	&-&\left.\frac{q}{2\pi}\int_0^\infty dy f_{\frac{D}{2}-1}\left(2r\cosh(y/2)\sqrt{m^2-\tilde{\mu}^2}\right)\frac{h(q,\alpha_0,y)}{\cosh(qy)-\cos(q\pi)}\right\}  \  .
\end{eqnarray}

\begin{figure}[tbph]
	\label{Fig_phi_1}
	\begin{center}
		\begin{tabular}{cc}
			\epsfig{figure=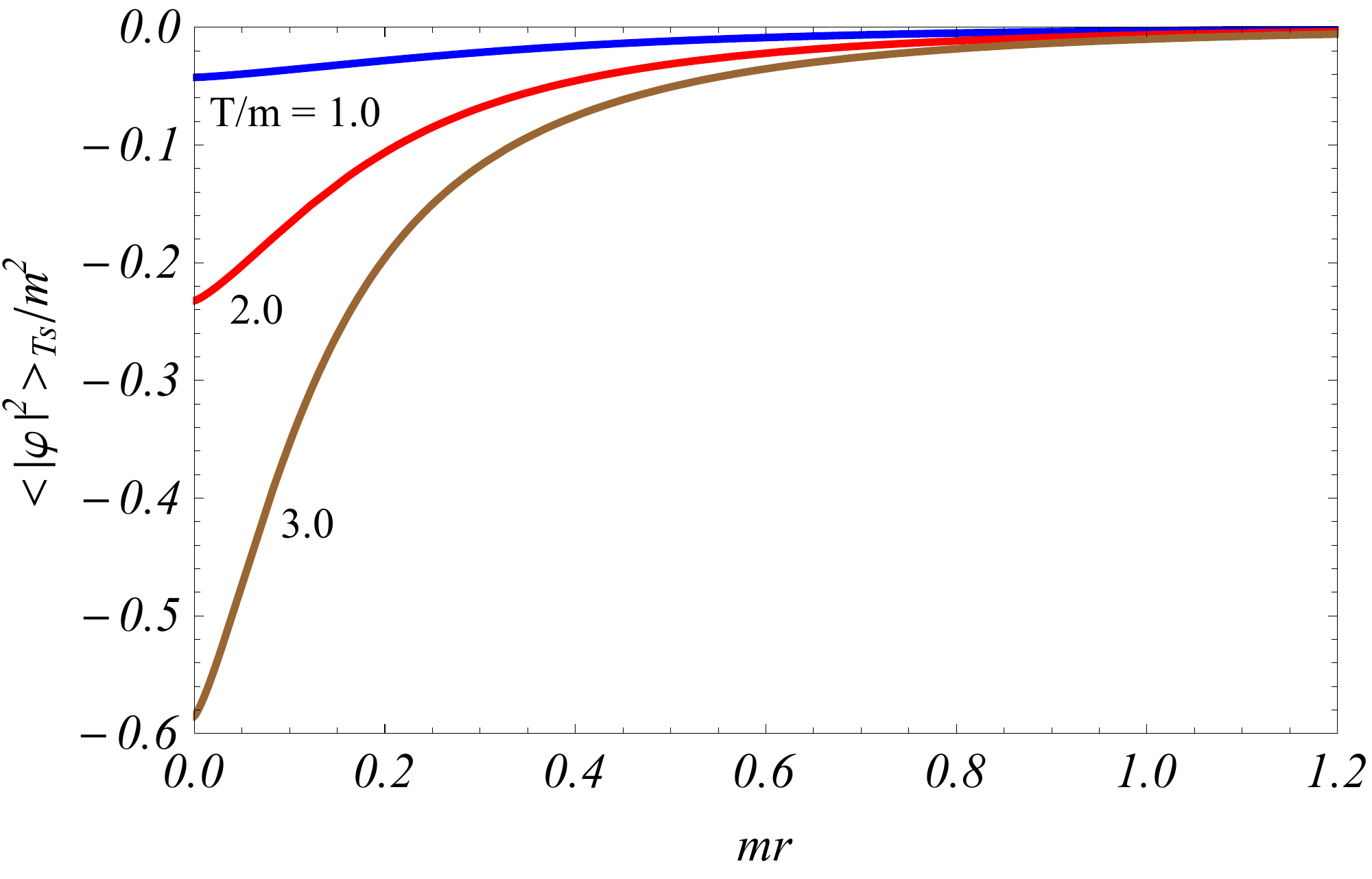,width=7.5cm,height=6cm} & \quad %
			\epsfig{figure=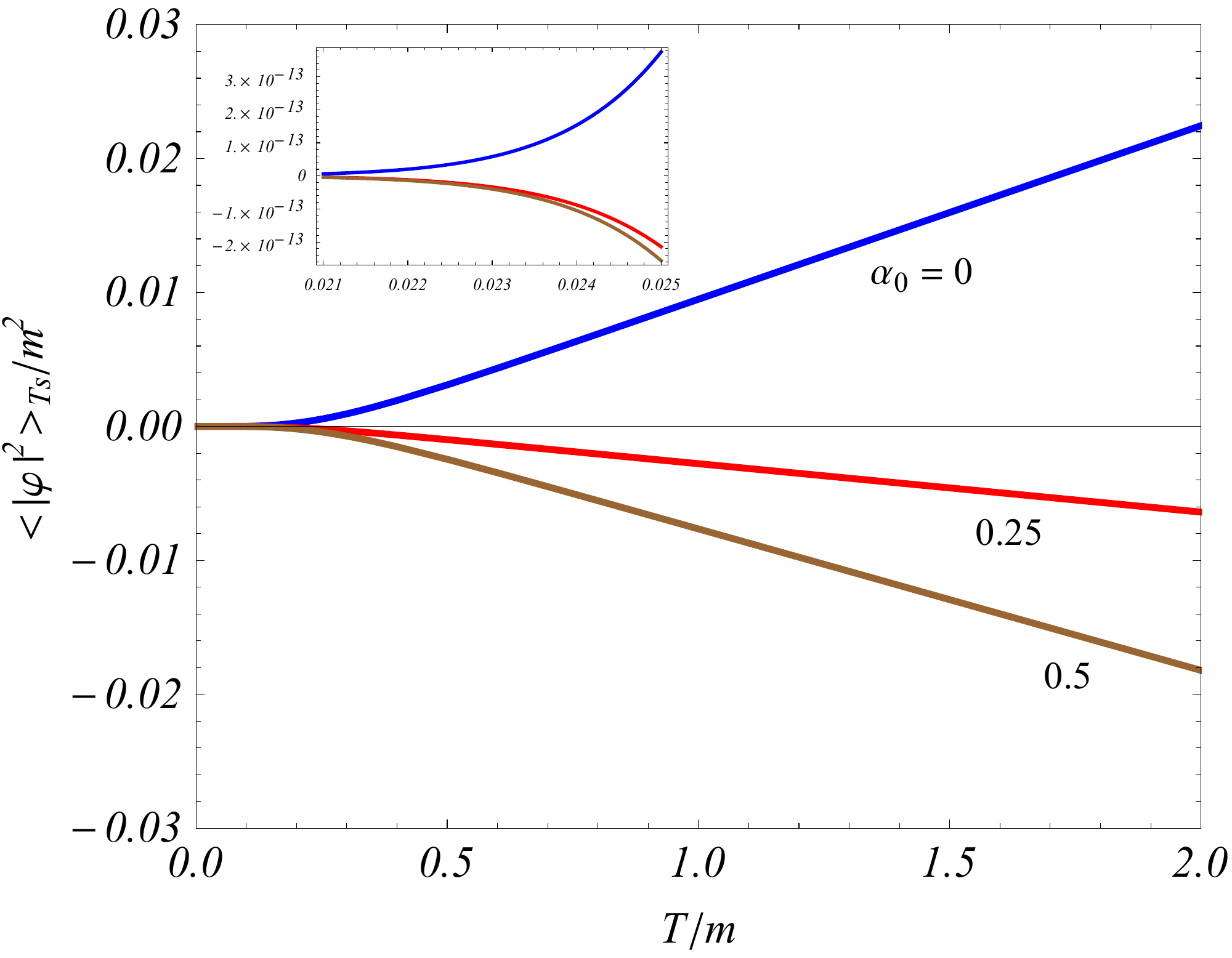,width=7.5cm,height=6cm}%
		\end{tabular}%
	\end{center}
	\caption{The thermal field squared induced by the string is plotted for $D=3$ as functions of the product $mr$ and the ratio $T/m$, with fixed parameters $q=2.5$, $|\tilde{\mu}|/m=0.5$ and $\alpha_{0}=0.25$. For the left and right panels the numbers near the curves correspond to different values of $T/m$ and $\alpha_0$, respectively.}
	\label{fig1}
\end{figure}

Considering $D=3$, in Fig. \ref{fig1} we exhibit the behavior $\langle|\varphi|^2\rangle_{Ts}$ as function of $mr$ for different values of $T/m$ and fixed $\alpha_0=0.25$ in the left pained, and as function $T/m$ for different values of $\alpha_0$ in the right panel with fixed $mr=1$. For both plots we take $q=2.5$ and $|\tilde{\mu}|/m=0.5$. From the right plot we can observe that depending the value of $\alpha_0$, the contribution of particle and antiparticle changes the intensity of $\langle|\varphi|^2\rangle_{Ts}$.

Our second analysis concerns to the contribution induced by the compactification. It is:
\begin{eqnarray}
	\label{phi2c}
	\langle|\varphi|^2\rangle_{Tc}&=&\frac{4m^{D-1}}{(2\pi)^{\frac{D+1}{2}}}\sum_{j=1}^\infty\cosh(j\beta\tilde{\mu})\sideset{}{'}\sum_{l=-\infty}^\infty e^{2\pi i l\tilde{\eta}}\left\{\sum_{k=1}^{[\frac q2]}\cos(2k\pi\alpha_0) f_{\frac{D-1}{2}}\left(m u_{jkl}\right)\right.\nonumber\\
	&-&\left.\frac{q}{2\pi}\int_0^\infty dy f_{\frac{D-1}{2}}\left(m u_{jyl}\right)\frac{h(q,\alpha_0,y)}{\cosh(qy)-\cos(q\pi)}\right\}  \  .
\end{eqnarray}
For this expression we have,
\begin{eqnarray}
	\label{u_1}
	u_{jkl}&=&\sqrt{4r^2\sin^2(k\pi/q)+(j\beta)^2+(lL)^2} \ , \nonumber\\
	u_{jyl}&=&\sqrt{4r^2\cosh^2(y/2)+(j\beta)^2+(lL)^2} \ .
\end{eqnarray}
Note that we have also omitted the term $k=0$ in the expression \eqref{phi2c}, which is a thermal contribution induced purely by the compactification in Minkowski spacetime: 
\begin{equation}
	\label{phi2_Min_c}
	\langle|\varphi|^2\rangle_{Tc}^{(M)}=\frac{2m^{D-1}}{(2\pi)^{\frac{D+1}{2}}}\sum_{j=1}^\infty\cosh(j\beta\tilde{\mu})\sum_{l=-\infty}^\infty e^{2\pi i l\tilde{\eta}}f_{\frac{D-1}{2}}(mu_{j0l}) \ .
\end{equation}
where $u_{j0l}=\sqrt{(j\beta)^2+(lL)^2}$. We also shall omit this contribution in our analysis below for the same reason given previously.

In the massless field case we have,
\begin{eqnarray}
	\label{phi2c0}
	\langle |\varphi|^2\rangle_{Tc}&=&\frac{\Gamma\left(\frac{D-1}{2}\right)}{\pi^{\frac{D+1}{2}}}\sum_{j=1}^{\infty}\sideset{}{'}\sum_{l=-\infty}^\infty e^{2\pi i l\tilde{\eta}}\Bigg\{\sum_{k=1}^{[\frac q2]}\frac{\cos(2k\pi\alpha_0)}{[4r^2\sin^2(\pi k/q)+(j\beta)^2+(lL)^2]^{(D-1)/2}} \nonumber \\
	&-&\frac{q}{2\pi}\int_0^\infty \frac{dy}{[4r^2\cosh^2(y/2)+(j\beta)^2+(lL)^2]^{(D-1)/2}}\frac{h(q,\alpha_0,y)}{\cosh(qy)-\cos(q\pi)}\Bigg\}  \  .
\end{eqnarray}

In the limit of large length of the extra dimension, $mL\gg1$, and considering $L\gg r$, the dominant contribution in \eqref{phi2c} comes from the terms with $j=1$, $l=-1$ and $l=1$:
\begin{eqnarray}
	\langle|\varphi|^2\rangle_{Tc}&\approx&\frac{4m^{D-1}g(q,\alpha_0)}{(2\pi)^{\frac{D}{2}}}\cosh(\beta\tilde{\mu}) \cos(2\pi  \tilde{\eta})\frac{e^{-\sqrt{(m\beta)^2+(mL)^2}}}{[(m\beta)^2+(mL)^2]^{D/4}}  \  ,
\end{eqnarray}
which goes to zero in the limit $L\rightarrow\infty$. 

At low temperature, $T\ll m, r^{-1}$, $\beta$ is large and once again using the asymptotic expansion of the Macdonald function for large arguments, one finds that in this regime the terms that most contribute are given for $j=1$ with $l=-1$ and with $l=1$:
\begin{eqnarray}
	\label{T_low1}
	\langle|\varphi|^2\rangle_{Tc}&\approx&\frac{2m^{\frac{D}{2}-1}g(q,\alpha_0)}{(2\pi)^{\frac{D}{2}}}\cos(2\pi\tilde{\eta})\frac{T^{D/2}e^{-(m-|\tilde{\mu}|)/T}}{[1+(TL)^2]^{D/4}}  \  ,
\end{eqnarray}
which also goes to zero at the zero-temperature limit. In both equations above,  
$g(q,\alpha_0)$ is given by \eqref{g-function}.

On the other side, at the high temperature limit, the leading contribution to \eqref{phi2c} comes from large $j$ and this representation is not appropriate to our evaluation. Once again we make use of the formula \eqref{Resum}, which in this case we identify $\alpha=i\beta\tilde{\mu}$,  $b={\small \begin{cases}
		\sqrt{4r^2\sin^2(\pi k/q)+(lL)^2} \\
		\sqrt{4r^2\cosh^2(y/2)+(lL)^2}
\end{cases}}$, $\nu=\frac{D-1}{2}$ and $w_{j}=\sqrt{(2\pi j/\beta+i\tilde{\mu})^2+m^2}$. Thus, the dominant term corresponds to $j=0$:
\begin{eqnarray}
	\label{T_high1}
	\langle|\varphi|^2\rangle_{Tc}&\approx&\frac{2(m^2-\tilde{\mu}^2)^{\frac{D}{2}-1}T}{(2\pi)^{\frac{D+1}{2}}}\sideset{}{'}\sum_{l=-\infty}^\infty e^{2\pi i l\tilde{\eta}}\left\{\sum_{k=1}^{[\frac q2]}\cos(2k\pi\alpha_0) f_{\frac{D}{2}-1}\left(u_{0kl}\sqrt{(m^2-\tilde{\mu}^2)}\right)\right.\nonumber\\
	&-&\left.\frac{q}{2\pi}\int_0^\infty dy f_{\frac{D}{2}-1}\left(u_{0yl}\sqrt{(m^2-\tilde{\mu}^2)}\right)\frac{h(q,\alpha_0,y)}{\cosh(qy)-\cos(q\pi)}\right\}  \  ,
\end{eqnarray}
with $u_{0kl}$ and $u_{0yl}$ are given in \eqref{u_1}.

\begin{figure}[tbph]
	\begin{center}
		\begin{tabular}{cc}
			\epsfig{figure=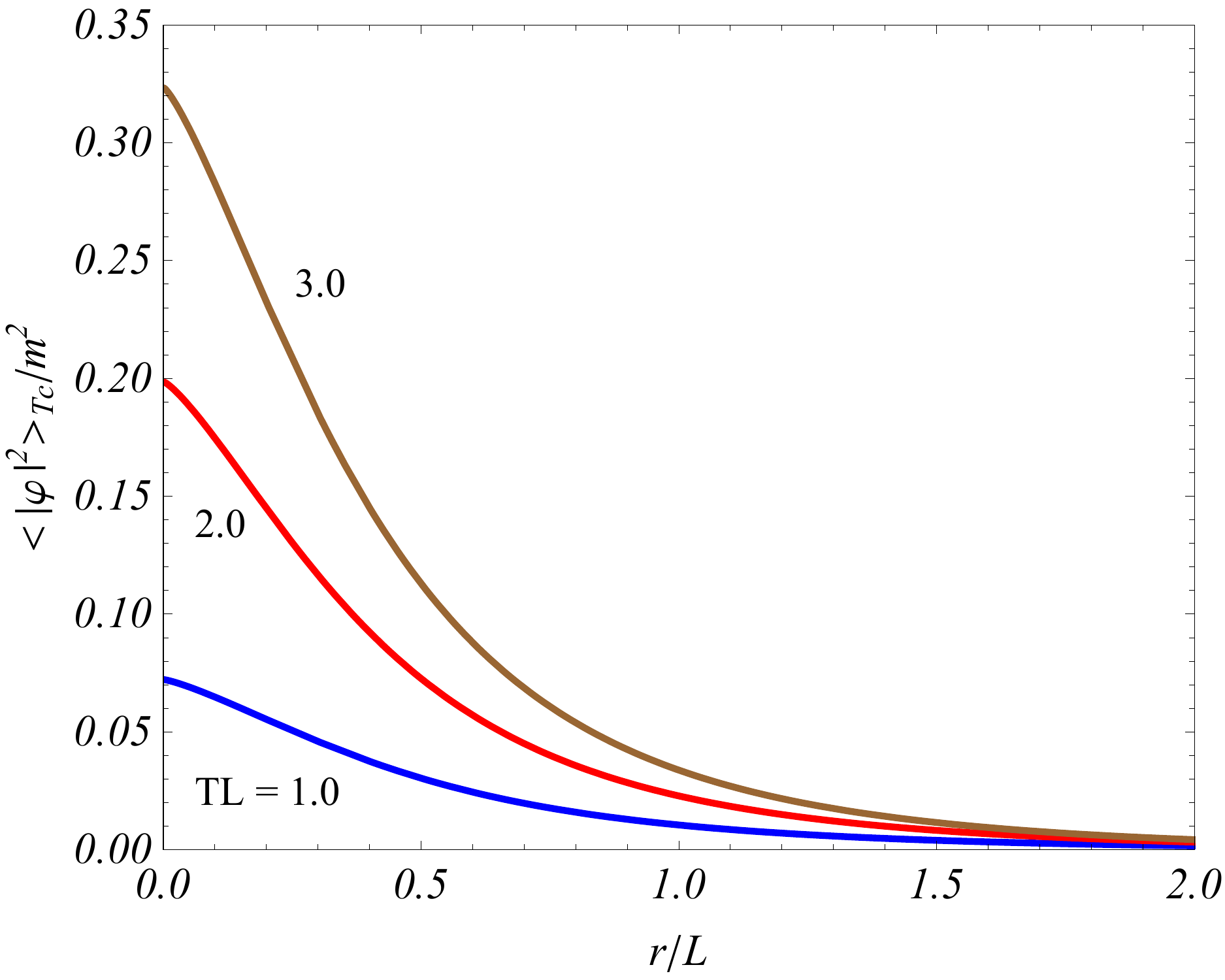,width=7.5cm,height=6cm} & \quad %
			\epsfig{figure=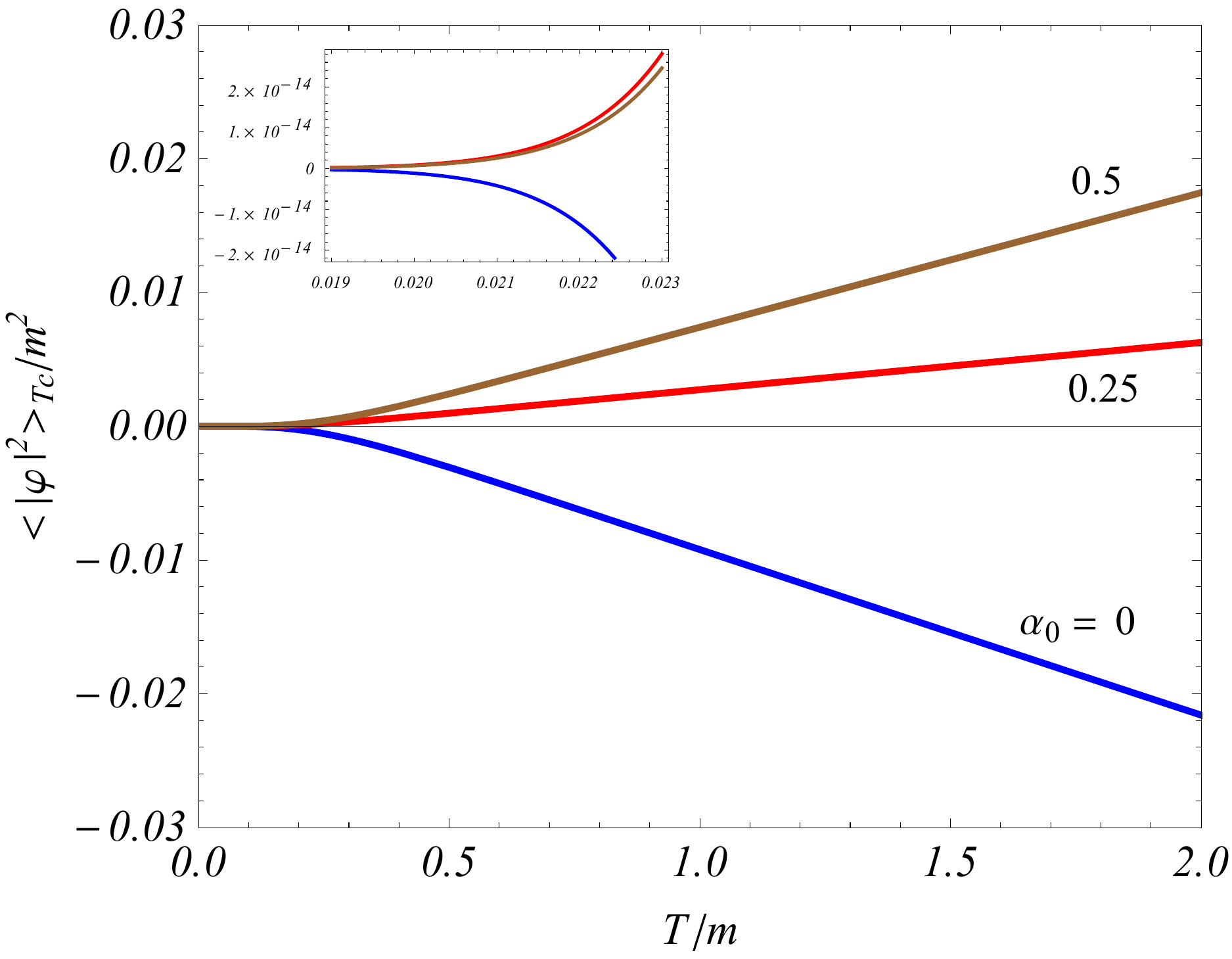,width=7.5cm,height=6cm}%
		\end{tabular}%
	\end{center}
	\caption{The thermal field squared induced by the compactification is plotted for $D=3$ as functions of the ratios $r/L$ and the $T/m$, with fixed parameters $q=2.5$, $|\tilde{\mu}|/m=0.5$, $\tilde{\eta}=0.5$ and $\alpha_{0}=0.25$. For the left panel we have also fixed $mr=1$ and the numbers near the curves correspond to different values of $TL$. For the right panel we have set $mL=0.75$ and the numbers near the curves correspond to different values of $\alpha_{0}$.}
	\label{fig2}
\end{figure}

Considering $D=3$ in Fig. \ref{fig2}, we exhibit in the left panel the behavior of the $\langle|\varphi|^2\rangle_{Tc}$ as function of $r/L$ for different values of $TL$ and fixing $mr=1$. In the right panel it is shown the behavior $\langle|\varphi|^2\rangle_{Tc}$ as function of $T/m$ for different values of $\alpha_0$ and the fixed parameters $mr=1$, $|\tilde{\mu}|/m=0.5$ and $mL=0.75$. For both plots we adopted $q=2.5$ and $\tilde{\eta}=0.5$.

Below, we present the total thermal correction to the field squared:
\begin{eqnarray}
	\label{phi_Total}
	\langle|\varphi|^2\rangle_{T}&=&\frac{4m^{D-1}}{(2\pi)^{\frac{D+1}{2}}}\sum_{j=1}^\infty\cosh(j\beta\tilde{\mu})\sum_{l=-\infty}^\infty e^{2\pi i l\tilde{\eta}}\left\{\sum_{k=1}^{[\frac q2]}\cos(2k\pi\alpha_0) f_{\frac{D-1}{2}}\left(m u_{jkl}\right)\right.\nonumber\\
	&-&\left.\frac{q}{2\pi}\int_0^\infty dy f_{\frac{D-1}{2}}\left(m u_{jyl}\right)\frac{h(q,\alpha_0,y)}{\cosh(qy)-\cos(q\pi)}\right\}  \  ,
\end{eqnarray}
where the component $l=0$ corresponds to the uncompactified contribution. Moreover, we use in the above expression the notation \eqref{u_1}

\section{Thermal  expectation value of the energy-momentum tensor}
\label{Energy_momentum}
One of the most important quantities which characterizes the properties of the quantum
vacuum is the VEV of the energy-momentum tensor. In addition to describing
the physical structure of the quantum field at a given point, the
energy-momentum tensor acts as a source of gravity in the Einstein equations.
For the system that we are considering, the VEV of energy-momentum tensor, has been investigated in \cite{Braganca_19}. In this section we are mainly interested to the thermal contribution to this quantity. In order to develop this calculation, we use the formula \cite{OliveiradosSantos:2019rjt},
\begin{equation}
	\langle T_{\mu\nu}\rangle_{T}=\frac{1}{2}(D_{\mu}D_{\nu'}^{\dagger}+D_{\mu'}^{\dagger}D_{\nu})G_{T}(x,x')-2[\xi R_{\mu\nu}+\xi\nabla_{\mu}\nabla_{\nu}-(\xi-1/4)g_{\mu\nu}\nabla_{\alpha}\nabla^{\alpha}]]\langle|\varphi|^2\rangle_{T},
	\label{EM-Tensor-formula}
\end{equation}
where the Ricci tensor, $R_{\mu\nu}$, vanishes for all points outside the string in Minkowski spacetime. In the above expression, we add the  factor $1/2$ in the first term of the right-hand side because we are using the thermal Hadamard function. 

Similar to what we have seen in the case of the field squared, the energy-momentum can be also decomposed as,
\begin{equation}
		\langle T_{\mu\nu}\rangle_T=	\langle T_{\mu\nu}\rangle_{Ts}+	\langle T_{\mu\nu}\rangle_{Tc} \ .
\end{equation}

Analyzing the contributions in the formula \eqref{EM-Tensor-formula}, we focus on the expectation values of the field squared \eqref{phi2s} and \eqref{phi2c}, whose theirs d'Alembertian of are given below:
\begin{eqnarray}
	\Box\langle |\varphi|^2\rangle_{Ts}&=&-\frac{32m^{D+1}}{(2\pi)^{\frac{D+1}{2}}}\Bigg\{\sideset{}{'}\sum_{k=1}^{[q/2]}\cos(2\pi k\alpha_0)\sin^2(\pi k/q)\left[2m^2r^2 s_k^2f_{\frac{D+3}{2}}(mu_{jk})-f_{\frac{D+1}{2}}(mu_{jk})\right]\nonumber\\
	&-&\frac{q}{\pi}\int_{0}^{\infty}dy\frac{\cosh^2(y/2)h(q,\alpha_0,y)}{\cosh(qy)-\cos(q\pi)}\left[2m^2r^2\cosh^2(y/2)f_{\frac{D+3}{2}}(mu_{jy})-f_{\frac{D+1}{2}}(mu_{jy})\right]\Bigg\} \nonumber\\
\end{eqnarray}
and
\begin{eqnarray}
	\Box\langle |\varphi|^2\rangle_{Tc}&=&-\frac{32m^{D+1}}{(2\pi)^{\frac{D+1}{2}}}\sum_{j=1}^\infty\cosh(j\beta\tilde{\mu})\sideset{}{'}\sum_{l=-\infty}^\infty e^{2\pi i l\tilde{\eta}}\Bigg\{\sideset{}{'}\sum_{k=1}^{[q/2]}\cos(2\pi k\alpha_0)\sin^2(\pi k/q)\nonumber\\
	&\times&\left[2m^2r^2s_k^2f_{\frac{D+3}{2}}(mu_{jkl})-f_{\frac{D+1}{2}}(mu_{jkl})\right]-\frac{q}{\pi}\int_{0}^{\infty}dy\frac{\cosh^2(y/2)h(q,\alpha_0,y)}{\cosh(qy)-\cos(q\pi)}\nonumber\\
	&\times&\left[2m^2r^2\cosh^2(y/2)lf_{\frac{D+3}{2}}(mu_{jyl})-f_{\frac{D+1}{2}}(mu_{jyl})\right]\Bigg\} \ ,
\end{eqnarray}
where in both equations above, we use the notation $s_k=\sin(\pi k/q)$.
In the geometry under consideration only the differential operators $\nabla_{r}\nabla_{r}$ and $\nabla_{\phi}\nabla_{\phi}$ produce non-vanishing terms when acting on the thermal expectation values of field squared.

The remaining contributions to the thermal expectation value of the  energy-momentum tensor come from the first term in the right-hand side of \eqref{EM-Tensor-formula}. In particular, for the operator $D_{\phi'}^{\dagger}D_{\phi}$, it is more convenient apply it in the representation \eqref{Thermal_s_1} of the Hadarmard function and subsequently take the coincidence limit in the angular variable. This procedure leads us to obtain the expression
\begin{equation}
	S(q,\alpha,\chi)=\sum_{n=-\infty}^{\infty}q^2(n+\alpha)^2I_{q|n+\alpha|}(\chi),
\end{equation}
where $\chi=urr'$. This sum can be developed by using the differential equation obeyed by the modified Bessel equation \cite{Grad}. Therefore, we get,
\begin{equation}
	S(q,\alpha,\chi)=\bigg(\chi^2\frac{d^2}{d\chi^2}+\chi\frac{d}{d\chi}-\chi^2\bigg)\sum_{n=-\infty}^{\infty}I_{q|n+\alpha|}(\chi),
\end{equation}
where this last sum is given by \cite{Braganca_15}
\begin{equation}
	\sum_{n=-\infty}^{\infty}I_{q|n+\alpha|}(\chi)=\frac{2}{q}\sideset{}{'}\sum_{k=0}^{[q/2]}\cos(2\pi k\alpha_0)e^{\chi\cos(2\pi k/q)}-\frac{1}{\pi}\int_{0}^{\infty}dy\frac{e^{-\chi\cosh(y)}f(q,\alpha_0,y)}{\cosh(qy)-\cos(q\pi)}.
\end{equation}

The contributions induced by the string without compactification and  by the compactified extra dimension, to the thermal average of the energy-momentum tensor is calculated from \eqref{EM-Tensor-formula}, by making use of the corresponding Wightman function and thermal average of the field
squared. After long but straightforward calculations, for the uncompactified and compactified contributions, one
finds (no summation over $\mu$):
\begin{eqnarray}
	\label{EM_s}
	\langle T_{\mu}^{\mu}\rangle_{Ts}&=&\frac{8m^{D+1}}{(2\pi)^{\frac{D+1}{2}}}\sum_{j=1}^{\infty}\cosh(j\beta\tilde{\mu})\Bigg[\sideset{}{'}\sum_{k=1}^{[q/2]}\cos(2\pi k\alpha_0)G_{\mu,0}^{\mu}(2mr,\sin(\pi k/q))\nonumber\\
	&-&\frac{q}{2\pi}\int_{0}^{\infty}dy\frac{h(q,\alpha_0,y)G_{\mu,0}^{\mu}(2mr,\cosh(y/2))}{\cosh(qy)-\cos(q\pi)}\Bigg],
	\label{EM-cosmic-string}
\end{eqnarray}
and
\begin{eqnarray}
	\label{EM_c}
	\langle T_{\mu}^{\mu}\rangle_{Tc}&=&\frac{8m^{D+1}}{(2\pi)^{\frac{D+1}{2}}}\sum_{j=1}^{\infty}\cosh(j\beta\tilde{\mu})\sideset{}{'}\sum_{l=-\infty}^\infty e^{2\pi i l\tilde{\eta}}\Bigg[\sideset{}{'}\sum_{k=1}^{[q/2]}\cos(2\pi k\alpha_0)G_{\mu,l}^{\mu}(2mr,\sin(\pi k/q))\nonumber\\
	&-&\frac{q}{2\pi}\int_{0}^{\infty}dy\frac{h(q,\alpha_0,y)G_{\mu,l}^{\mu}(2mr,\cosh(y/2))}{\cosh(qy)-\cos(q\pi)}\Bigg],
	\label{EM-compactification}
\end{eqnarray}
where
\begin{eqnarray}
	G_{0,l}^{0}(u,v)&=&(mj\beta)^2f_{\frac{D+3}{2}}(w_l)+(1-4\xi)v^2[(uv)^2f_{\frac{D+3}{2}}(w_l)-2f_{\frac{D+1}{2}}(w_l)	]-f_{\frac{D+1}{2}}(w_l)\nonumber\\
	G_{1,l}^{1}(u,v)&=&(4\xi v^2-1)f_{\frac{D+1}{2}}(w_l)\nonumber\\
	G_{2,l}^{2}(u,v)&=&(1-4\xi v^2)[(uv)^2f_{\frac{D+3}{2}}(w_l)-f_{\frac{D+1}{2}}(w_l)	]\nonumber\\
	G_{3,l}^{3}(u,v)&=&(1-4\xi)v^2[(uv)^2f_{\frac{D+3}{2}}(w_l)-2f_{\frac{D+1}{2}}(w_l)	]-f_{\frac{D+1}{2}}(w_l)\nonumber\\
	&+&(mlL)^2f_{\frac{D+3}{2}}(w_l) \ ,
	\nonumber\\
	G_{i,l}^{i}(u,v)&=&(1-4\xi)v^2[(uv)^2f_{\frac{D+3}{2}}(w_l)-2f_{\frac{D+1}{2}}(w_l)	]-f_{\frac{D+1}{2}}(w_l) \ , \ \text{with $i=4,...,D.$}
	\label{G-functions}
\end{eqnarray}
with
\begin{equation}
	w_l=\sqrt{(uv)^2+(mj\beta)^2+(mlL)^2} \ .
	\label{arg_wl}
\end{equation}
In the expressions above we have omitted the contribution coming from Minkowski spacetime, that correspond to the $k=0$ component on the summation. The reason is the same as  explained in the last section. We are only interested in the analysis of the thermal correction on the energy-momentum tensor induced by the compactified cosmic string spacetime.
 
 At this point we want to call attention that the presence of the term $(mj\beta)^2$ in $G_{0,l}^{0}(u,v)$, has a similar structure of the term $(mlL)^2$ in $G_{3,l}^{3}(u,v)$. In fact, this similarity can be understood through the Matsubara formalism where, in order to treat a system with non-zero temperature, one uses the Euclidian time (obtained  through a rotation of the time coordinate, $t\rightarrow i\tau$) and confines it to the interval $\tau\in[0,\beta]$, with $\beta=1/T$ \cite{Bordag:2009}. In this case, $\beta$ is the analogue of the spatial compactification length, $L$, and the scalar field must obey a periodic condition, $\varphi(\tau+\beta,\vec{x})=\varphi(\tau,\vec{x})$, which is the analogue of the more general spatial quasi-periodic boundary condition \eqref{QPC}.

Let us now analyse some limiting cases for the energy density component, $\langle T_{0}^{0}\rangle_{T}$, for the uncompactified and compactified induced contributions. We start analysing the cosmic string induced contribution. 

For a conformal coupled massless scalar field, the energy density component induced by the string reads,
\begin{eqnarray}
	\langle T_{0}^{0}\rangle_{Ts}&=&\frac{8\Gamma\left(\frac{D+1}{2}\right)}{\pi^{\frac{D+1}{2}}}\sum_{j=1}^{\infty}\Biggl\{\sideset{}{'}\sum_{k=1}^{[q/2]}\frac{\cos(2\pi k\alpha_0)}{u_{jk}^{D+1}}\Bigg[\frac{D+1}{2}\nonumber\\
	&\times&\left(\frac{j^2\beta^2+4r^2\sin^4(\pi k/q)/D}{u_{jk}^2}\right)
	-\left(\frac{\sin^2(\pi k/q)}{D}+\frac{1}{2}\right)\Bigg]
	\nonumber\\
	&-&\frac{q}{2\pi}\int_{0}^{\infty}dy\frac{h(q,\alpha_0,y)}{[\cosh(qy)-\cos(q\pi)]u_{jy}^{D+1}}\Bigg[\frac{D+1}{2}\nonumber\\
	&\times&\left(\frac{j^2\beta^2+4r^2\cosh^4(y/2)/D}{u_{jy}^2}\right)
	-\left(\frac{\cosh^2(y/2)}{D}+\frac{1}{2}\right)\Bigg]\Biggr\} \ .
\end{eqnarray}

Now we want to analyze the asymptotic behavior of the thermal correction to the energy density component
associated with the uncompactified cosmic string, $\langle T_{0}^{0}\rangle_{Ts}$, for points near the string's core. In the absence of the magnetic flux running along the string, this term is finite at $r=0$; however the presence of the magnetic flux change the situation. The main reason is due to the expression for the function $h(q,\alpha_{0},y)$ given in \eqref{h}. As we can see, taking $\alpha_{0}=0$, there is no hyperbolic cosines involving the variable $y$. In addition to this function, the integrand also presents the term $G_{0,0}^{0}$ given in \eqref{G-functions}. In order to analyze the finiteness of the integral, we have to observe its integrand for large values of $y$. In this limit, we can approximate $\cosh(y/2)\approx e^{y/2}/2$. So, taking this procedure in consideration, we conclude that the integral is finite for $|\alpha_{0}|>1/q$. For this case, we can put directly $r=0$ in the function $G_{0,0}^{0}$. For $|\alpha_{0}|<1/q$ the integral is divergent. A careful analysis, by considering the approximation $\cosh(y/2)\approx e^{y/2}/2$, allow us to conclude that the energy density induced by the string diverges like $1/(mr)^{2(1-q|\alpha_{0}|)}$ on the string's core.

\begin{figure}[tbph]
	\begin{center}
		\begin{tabular}{cc}
			\epsfig{figure=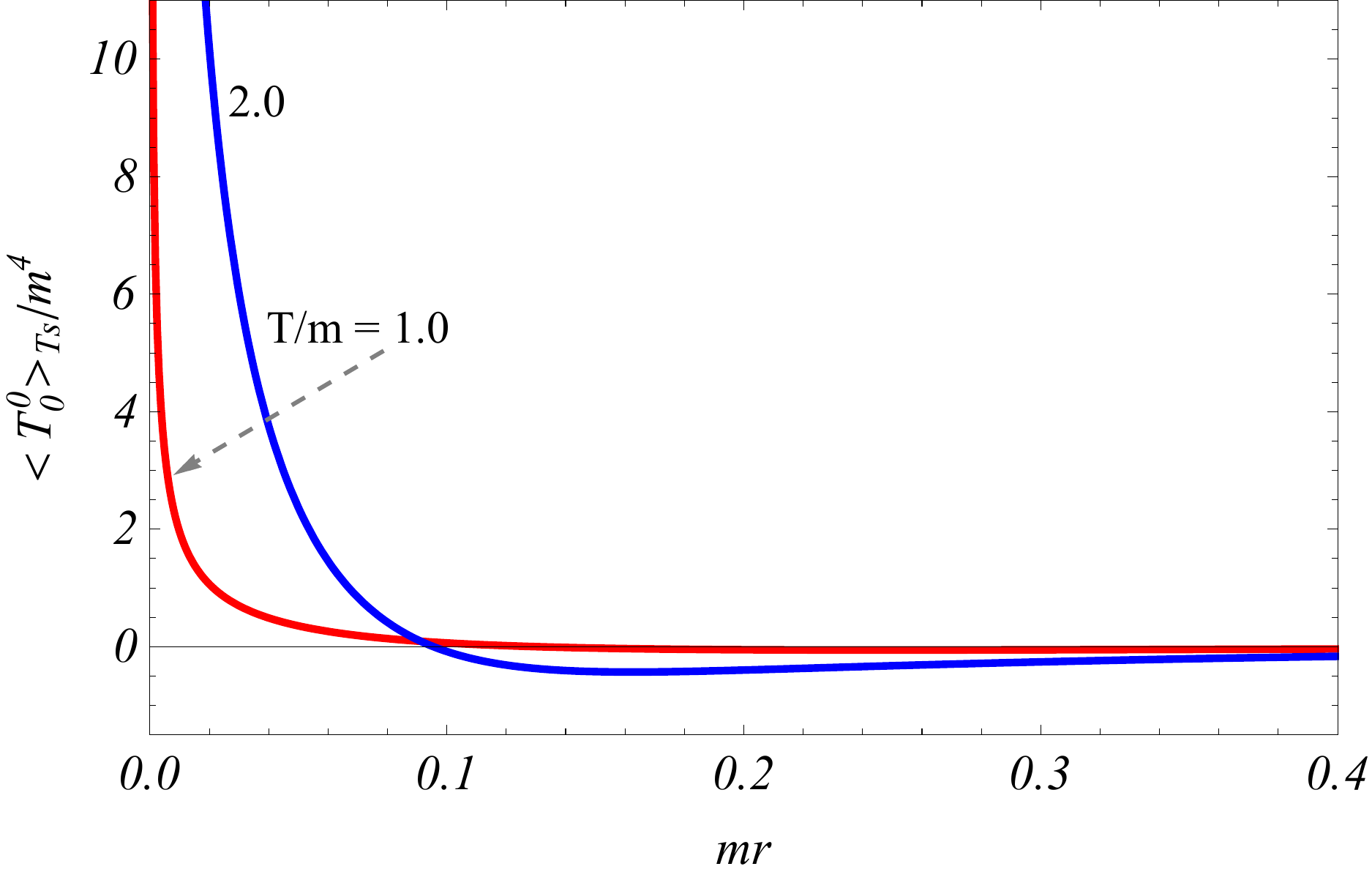,width=7.5cm,height=6cm} & \quad %
			\epsfig{figure=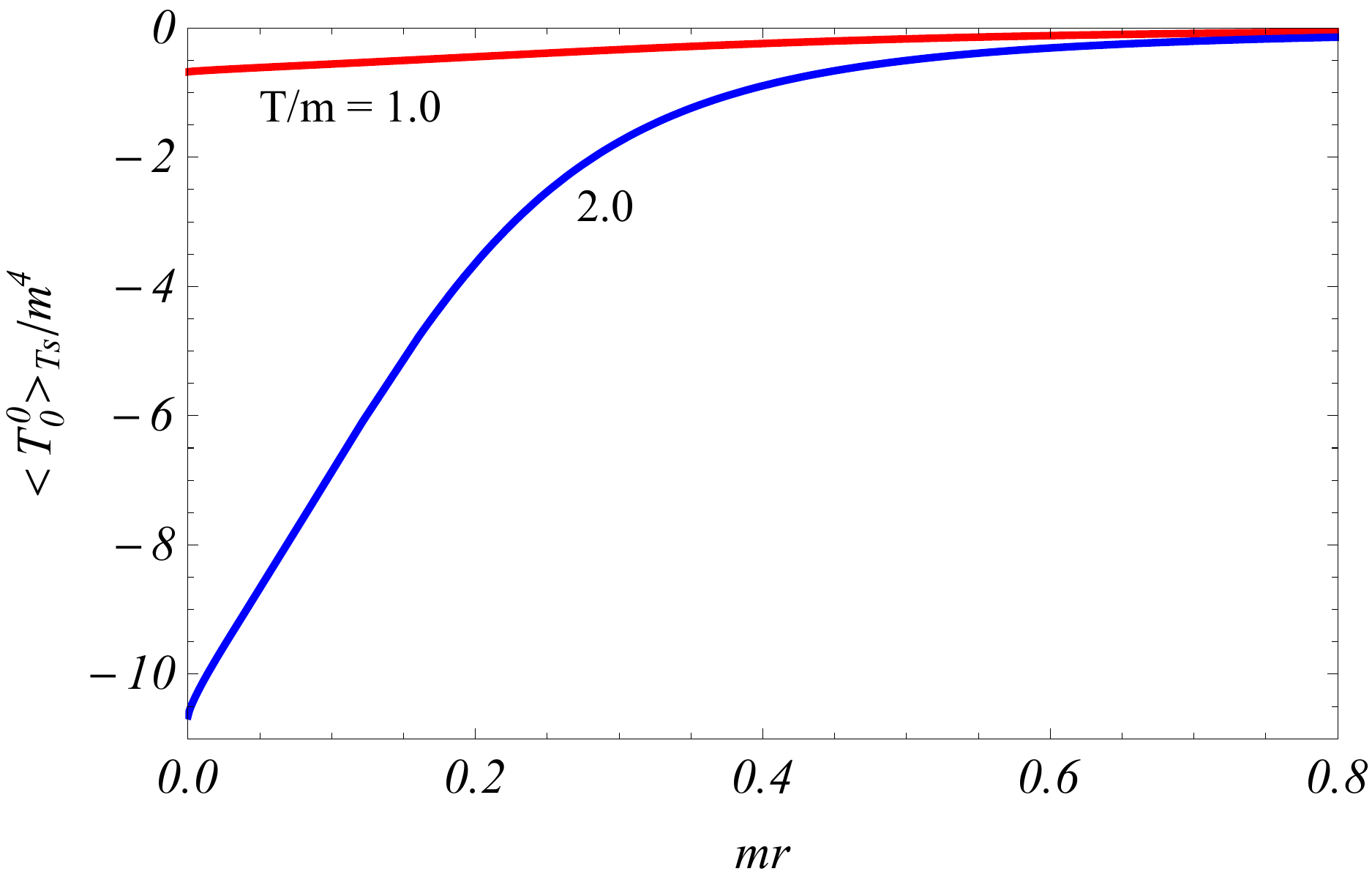,width=7.5cm,height=6cm}%
		\end{tabular}
	\end{center}
	\caption{The $\langle T_{0}^{0}\rangle_{Ts}$ part is plotted for $D=3$ as a function of the product $mr$, with fixed parameters $q=3.5$, $|\tilde{\mu}|/m=0.5$ and $\xi=0$. The left panel is plotted for $\alpha_{0}=0.2$ and the right panel for $\alpha_{0}=0.4$. The numbers near the curves correspond to values of $T/m$.}
		\label{Fig3}
\end{figure}

In Fig. \ref{Fig3} we plot the $\langle T_{0}^{0}\rangle_{Ts}$ for $D=3$, considering the fixed parameters $q=3.5$, $|\tilde{\mu}|/m=0.5$ and $\xi=0$ (minimal coupling) as a function of the product $mr$, with different values of the ratio $T/m$ and $\alpha_{0}$. The left panel is plotted for $\alpha_{0}=0.2$ and the right panel for $\alpha_{0}=0.4$. The numbers near the curves correspond to values of $T/m$. In correspondence to the asymptotic analysis made above, the curves with $\alpha_{0}=0.2$ satisfy the constraint relation $q|\alpha_{0}|<1$ and therefore diverge on the string's core, $r=0$, while the curves for $\alpha_{0}=0.4$ obey the constraint $q|\alpha_{0}|>1$ and are finite on the string.

Let us now analyze the limit $mr\gg1$. For $q>2$, the leading order term is given by the term $k=1$ and $j=1$:
\begin{eqnarray}
	\langle T_{0}^{0}\rangle_{Ts}\approx\frac{8(1-4\xi)m^{\frac{D}{2}+1}e^{\beta|\tilde{\mu}|}}{(2\pi)^{\frac{D}{2}}r^{D/2}}\cos(2\pi \alpha_0)\sin^4(\pi /q)\frac{e^{-m\sqrt{4r^2\sin^2(\pi /q)+\beta^2}}}{[4\sin^2(\pi /q)+\beta^2/r^2]^{\frac{D}{4}+1}} \ .
\end{eqnarray}
On the other hand, for $1\le q<2$, the dominant contribution is given by,
\begin{eqnarray}
	\langle T_{0}^{0}\rangle_{Ts}&\approx&-\frac{8(1-4\xi)qm^{\frac{D}{2}+1}e^{\beta|\tilde{\mu}|}}{(2\pi)^{\frac{D}{2}+1}r^{D/2}}\int_{0}^{\infty}dy\frac{h(q,\alpha_0,y)\cosh^4(y/2)}{\cosh(qy)-\cos(q\pi)}\nonumber\\
	&\times&\frac{e^{-m\sqrt{4r^2\cosh^2(y/2)+\beta^2}}}{[4\cosh^2(y/2)+\beta^2/r^2]^{\frac{D}{4}+1}} \ .
\end{eqnarray}

Similar to the field squared, at the low temperature limit, $T\ll m, r^{-1}$, the parameter $\beta$ is large and we can again use the corresponding asymptotic expansion for the function $f_{\nu}(x)$ in this limiting case. The leading order contribution in this case comes from the term with $j=1$:
\begin{equation}
	\label{EM_Tlow_s}
	\langle T_{0}^{0}\rangle_{Ts}\approx\frac{2m^{\frac{D}{2}+1}g(q,\alpha_{0})}{(2\pi)^{\frac{D}{2}}}T^{D/2}e^{-(m-|\tilde{\mu}|)/T} \ .
\end{equation}

On the other hand, at high temperature limit, $T\gg m,r^{-1}$, the use of the \eqref{Resum} is necessary again, because the dominant contribution comes from large $j$. In the present case, we have the identifications $\alpha=i\beta\tilde{\mu}$,  $b=2r{\small \begin{cases}
		\sin(\pi k/q) \\
		\cosh(y/2)
\end{cases}}$ and $w_{j}=\sqrt{(2\pi j/\beta+i\tilde{\mu})^2+m^2}$. After some straightforward intermediate steps, we observe that the leading contribution comes from the term $j=0$:
\begin{eqnarray}
	\langle T_{0}^{0}\rangle_{Ts}&\approx&\frac{4T}{(2\pi)^{\frac{D}{2}}}\Bigg[\sideset{}{'}\sum_{k=1}^{[q/2]}\cos(2\pi k\alpha_0)\tilde{g}(2mr,\sin(\pi k/q))\nonumber\\
	&-&\frac{q}{2\pi}\int_{0}^{\infty}dy\frac{h(q,\alpha_0,y)\tilde{g}(2mr,\cosh(y/2))}{\cosh(qy)-\cos(q\pi)}\Bigg],
	\label{EM-cs-High-Temp}
\end{eqnarray}
where we have defined the function,
\begin{eqnarray}
	\tilde{g}(u,v)&=&2\left(1+\frac{D}{2}\right)\left(\frac{D\tilde{\mu}^2}{m^2-\tilde{\mu}^2}-1\right)(m^2-\tilde{\mu}^2)^{D/2}f_{D/2+1}(u\sqrt{m^2-\tilde{\mu}^2})\nonumber\\
	&+&2\left(D+\frac{3}{2}\right)u\tilde{\mu}^2(m^2-\tilde{\mu}^2)^{(D-1)/2}f_{D/2+2}(u\sqrt{m^2-\tilde{\mu}})\nonumber\\
	&-&u(m^2-\tilde{\mu}^2)^{(D+1)/2}f_{D/2+2}(u\sqrt{m^2-\tilde{\mu}^2})+\tilde{\mu}^2u^2(m^2-\tilde{\mu}^2)^{D/2}f_{D/2+3}(u\sqrt{m^2-\tilde{\mu}^2})\nonumber\\
	&+&(1-4\xi)u^2v^4(m^2-\tilde{\mu}^2)^{D/2+1}f_{D/2+1}(u\sqrt{m^2-\tilde{\mu}^2})\nonumber\\
	&-&[1+2(1-4\xi)v^2](m^2-\tilde{\mu}^2)^{D/2}f_{D/2}(u\sqrt{m^2-\tilde{\mu}^2}) \ .
	\label{g-func}
\end{eqnarray}

\begin{figure}[tbph]
	\label{Fig_T00cs_TperM}
	\begin{center}
		\begin{tabular}{cc}
			\epsfig{figure=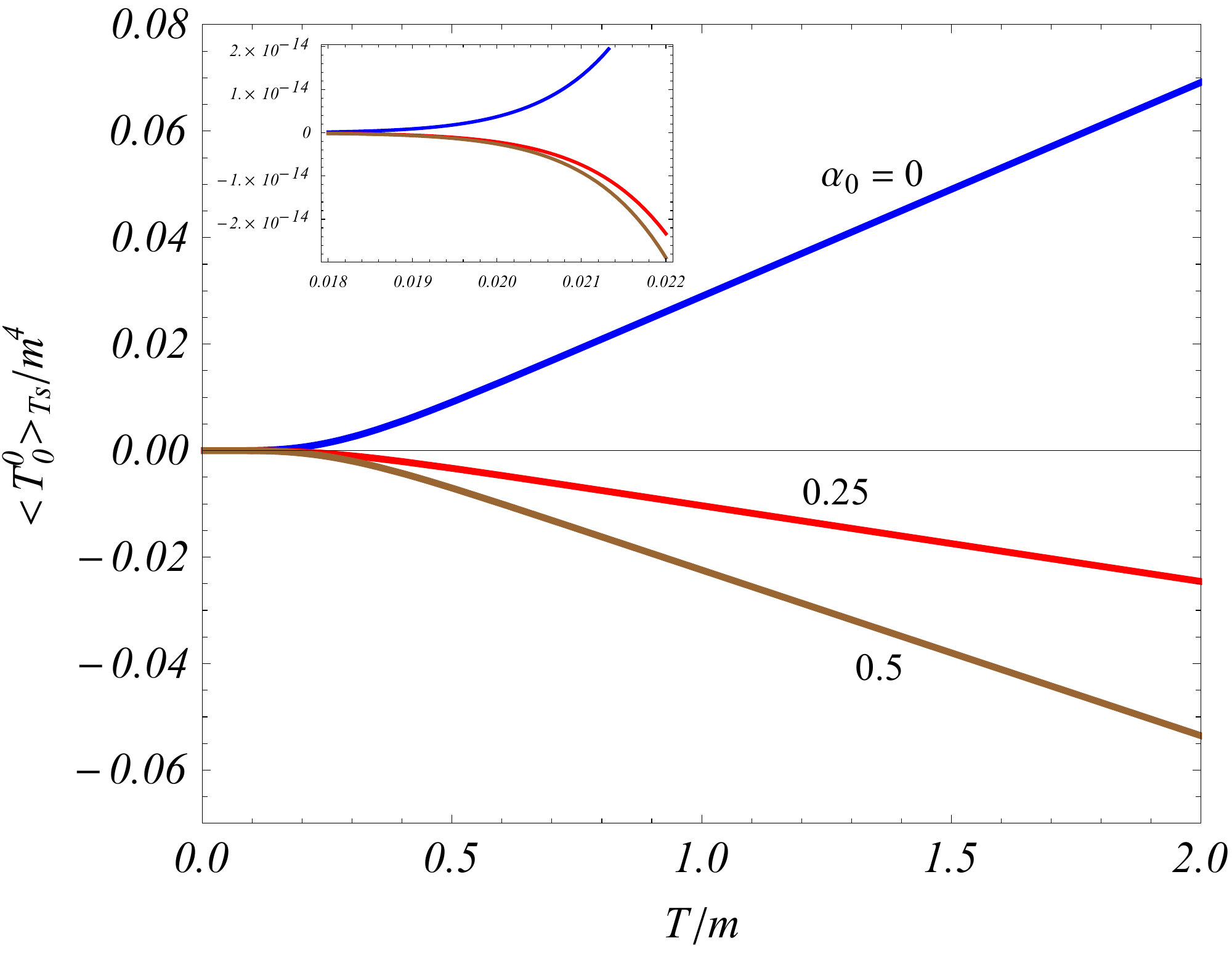,width=7.5cm,height=6cm}%
		\end{tabular}%
	\end{center}
\caption{The thermal energy density contribution induced by the string is plotted for $D=3$ as a function of the ratio $T/m$, with fixed parameters $q=2.5$, $|\tilde{\mu}|/m=0.5$, $mr=1$ and $\xi=0$. Numbers near the curves correspond to different values of $\alpha_0$.}
\end{figure}

In Fig. \ref{Fig_T00cs_TperM} we display the dependence of the thermal  energy density induced by the string as function of  ratio $T/m$, considering $D=3$ for $q=2.5$, $|\tilde{\mu}|/m=0.5$, $mr=1$, $\xi=0$ and different values of $\alpha_{0}$. This plot confirms the asymptotic behaviours analyzed above for both low and high temperature regimes.

Now, the contribution induced by the compactification in the conformally coupled massless field case has the form,
\begin{eqnarray}
	\langle T_{0}^{0}\rangle_{Tc}&=&\frac{8\Gamma\left(\frac{D+1}{2}\right)}{\pi^{\frac{D+1}{2}}}\sum_{j=1}^{\infty}\sideset{}{'}\sum_{l=-\infty}^\infty e^{2\pi i l\tilde{\eta}}\Biggl\{\sideset{}{'}\sum_{k=1}^{[q/2]}\frac{\cos(2\pi k\alpha_0)}{u_{jkl}^{D+1}}\Bigg[\frac{D+1}{2}\nonumber\\
	&\times&\left(\frac{j^2\beta^2+4r^2\sin^4(\pi k/q)/D}{u_{jkl}^2}\right)
	-\left(\frac{\sin^2(\pi k/q)}{D}+\frac{1}{2}\right)\Bigg]
	\nonumber\\
	&-&\frac{q}{2\pi}\int_{0}^{\infty}dy\frac{h(q,\alpha_0,y)}{[\cosh(qy)-\cos(q\pi)]u_{jyl}^{D+1}}\Bigg[\frac{D+1}{2}\nonumber\\
	&\times&\left(\frac{j^2\beta^2+4r^2\cosh^4(y/2)/D}{u_{jyl}^2}\right)
	-\left(\frac{\cosh^2(y/2)}{D}+\frac{1}{2}\right)\Bigg]\Biggr\} \ .
\end{eqnarray}

For large lengths of the extra dimension, $Lm\gg1$, and $r\ll L$, the dominant contribution comes from the terms for $j=1$ and $l=-1$ and for $j=1$ and $l=1$, and it is written as,
\begin{eqnarray}
	\langle T_{0}^{0}\rangle_{Tc}&\approx&-\frac{8m^{D+1}\cosh(\beta\tilde{\mu})\cos(2\pi \tilde{\eta})}{(2\pi)^{\frac{D}{2}}}\frac{e^{-m\sqrt{\beta^2+L^2}}}{[\beta^2+L^2]^{\frac{D+2}{4}}}\Biggl\{\sideset{}{'}\sum_{k=1}^{[q/2]}\cos(2\pi k\alpha_0)\nonumber\\
	&\times&[2(1-4\xi)\sin^2(\pi k/q)+1]-\frac{q}{2\pi}\int_{0}^{\infty}dy\frac{h(q,\alpha_0,y)}{\cosh(qy)-\cos(q\pi)}\nonumber\\
	&\times&[2(1-4\xi)\cosh^2(y/2)+1]\Biggr\} \ ,
\end{eqnarray}
which is valid only in the region $|\alpha_{0}|>1/q$. On the other hand, for $|\alpha_{0}|<1/q$ the integral term diverges at the upper limit. This divergence comes from the fact that we cannot ignore $4m^2r^2\cosh^2(y/2)$ in \eqref{arg_wl} with respect to $(mj\beta)^2+(mlL)^2$ in the integral term, i.e., we cannot disregard the behaviour of $\langle T_{0}^{0}\rangle_{Tc}$ near the string's core, $r=0$, when considering the region of parameters $|\alpha_{0}|<1/q$. Following a similar procedure adopted for the analysis of the behaviour of $\langle T_{0}^{0}\rangle_{Ts}$ near the string's core made above, one can show that the compactification induced term also diverges like $1/(mr)^{2(1-q|\alpha_{0}|)}$ on the string's core.

Now considering the energy density induced by compactification at the low temperature limit, $T\ll m, r^{-1}$, we make use once more of the corresponding asymptotic expansion for the function $f_{\nu}(x)$ in this limiting case, since the parameter $\beta$ is large. The leading order contribution in this case comes from the terms for $j=1$ and $l=-1$ and for $j=1$ and
 $l=1$:
\begin{equation}
	\label{EM_Tlow_c}
	\langle T_{0}^{0}\rangle_{Tc}\approx\frac{4m^{\frac{D}{2}+1}g(q,\alpha_{0})\cos(2\pi \tilde{\eta})}{(2\pi)^{\frac{D}{2}}}\frac{T^{D/2}e^{\big(|\tilde{\mu}|-m\sqrt{1+(TL)^2}\big)/T}}{[1+(TL)^2]^{\frac{D}{4}+1}} \ .
\end{equation}

On the other side, at high temperature limit, $T\gg m,r^{-1}$, we use again the \eqref{Resum}, with $\alpha=i\beta\tilde{\mu}$,  $b=2r{\small \begin{cases}
			\sqrt{4r^2\sin^2(\pi k/q)+(lL)^2} \\
		\sqrt{4r^2\cosh^2(y/2)+(lL)^2}
	\end{cases}}$ and $w_{j}=\sqrt{(2\pi j/\beta+i\tilde{\mu})^2+m^2}$. The dominant contribution also in this case comes from the term $j=0$:
\begin{eqnarray}
	\langle T_{0}^{0}\rangle_{Tc}&\approx&\frac{4T}{(2\pi)^{\frac{D}{2}}}\sum_{l=-\infty}^\infty e^{2\pi i l\tilde{\eta}}\Bigg[\sideset{}{'}\sum_{k=1}^{[q/2]}\cos(2\pi k\alpha_0)\tilde{g}\left(	\sqrt{4r^2\sin^2(\pi k/q)+(lL)^2},\sin(\pi k/q)\right)\nonumber\\
	&-&\frac{q}{2\pi}\int_{0}^{\infty}dy\frac{h(q,\alpha_0,y)\tilde{g}\left(\sqrt{4r^2\cosh^2(y/2)+(lL)^2},\cosh(y/2)\right)}{\cosh(qy)-\cos(q\pi)}\Bigg],
	\label{EM-c-High-Temp}
\end{eqnarray}
where the function $\tilde{g}(u,v)$ is defined in \eqref{g-func}.

\begin{figure}[tbph]
	\begin{center}
		\begin{tabular}{cc}
			\epsfig{figure=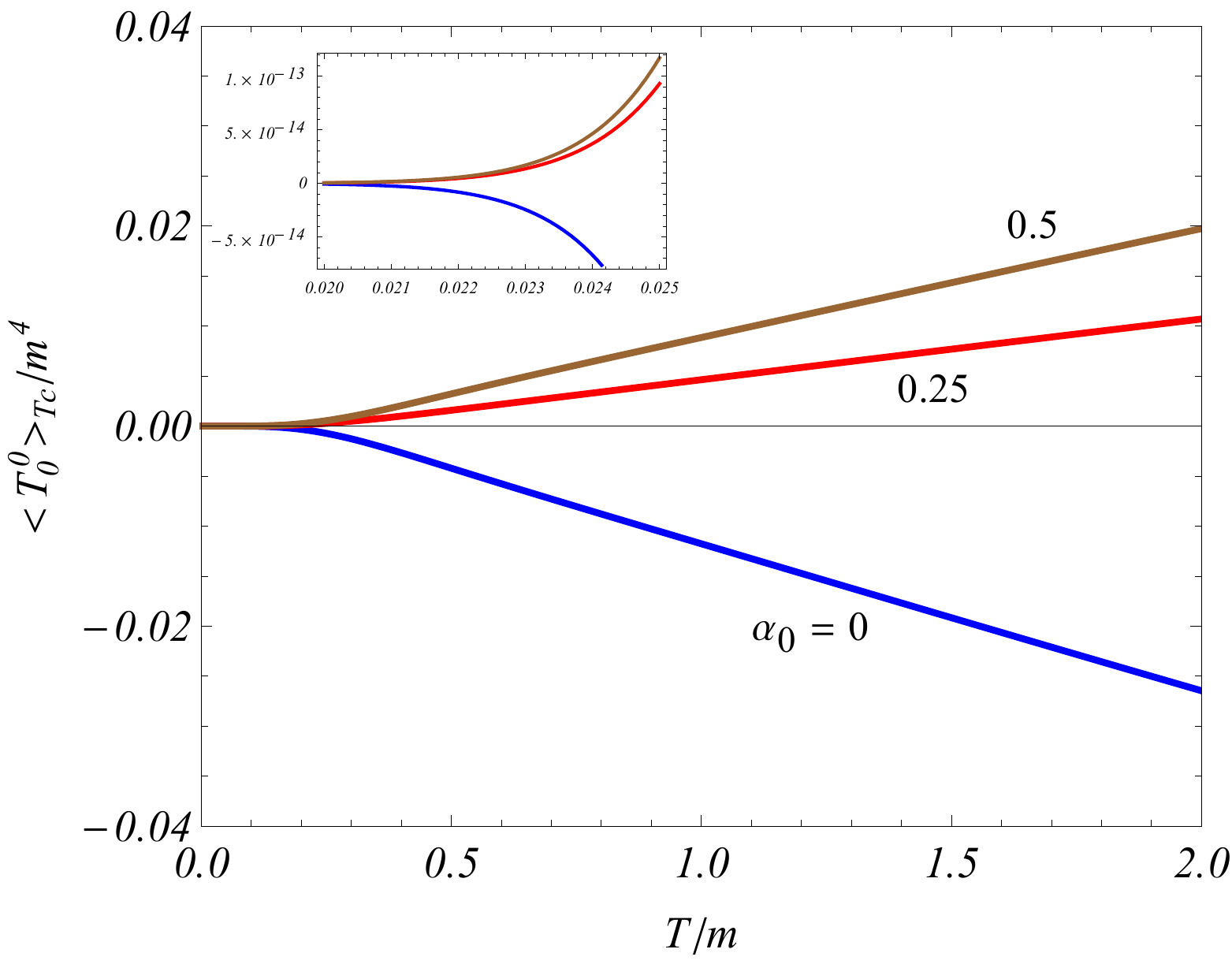,width=7.5cm,height=6cm}%
		\end{tabular}%
	\end{center}
	\caption{The energy density contribution induced by the compactification is plotted for $D=3$ as a function of the ratio $T/m$, with fixed parameters $q=2.5$, $mr=1$, $L/m=1$, $|\tilde{\mu}|/m=0.5$, $\tilde{\eta}=0.5$ and $\xi=0$. The numbers near the curves represent the different values of $\alpha_0$.}
	\label{Fig_T00c_TperM}
\end{figure}

The Fig. \ref{Fig_T00c_TperM} presents the $\langle T_{0}^{0}\rangle_{Tc}$ contribution in the thermal energy density as function of the ratio $T/m$, considering $D=3$, for fixed $q=2.5$, $mr=1$, $L/m=1$, $\xi=0$, $|\tilde{\mu}|/m=0.5$, $\tilde{\eta}=0.5$ and different values of $\alpha_{0}$.

As a last verification step of the results found in our investigation, one can check that the components of the energy-momentum tensor for both uncompactified and compactified induced contributions satisfy the covariant conservation equation, $\nabla_{\mu}\langle T_{\nu}^{\mu}\rangle=0$. For the spacetime geometry under consideration, this equation is reduced to $\langle T_{\phi}^{\phi}\rangle=\partial_{r}(r\langle T_{r}^{r}\rangle)$. On top of that, as second verification step, one can also verify that they satisfy the trace relation,
\begin{equation}
	\langle T_{\mu}^{\mu}\rangle=2\left[D(\xi-\xi_{D})\nabla_{\mu}\nabla^{\mu}\langle |\varphi|^2\rangle+m^2\langle |\varphi|^2\rangle\right] \ ,
\end{equation}
which is zero for the conformally coupled massless quantum scalar field case.
\section{Conclusions}
\label{conc} 
In this paper, we have investigated the finite temperature expectation values of the field squared and the energy-momentum tensor density for a massive bosonic quantum field with nonzero chemical potential in the geometry of a higher dimensional compactified cosmic string spacetime containing magnetic fluxes, one along the string's core and the other enclosed by the compact dimension. In contrast to the fermionic chemical potential which in general can have any value, the bosonic one is restricted by $|\tilde{\mu}|\leqslant E_0$, being  $E_0$  the minimum of energy. In order to calculate the thermal expectation value of these densities at temperature $T$, we had to obtain the thermal Hadamard function. Working with the grand canonical ensemble and also expanding the field operator in terms of a complete set of normalized positive and negative energy solutions of Klein-Gordon equation, we have decomposed the Hadamard function and consequently the densities to the vacuum expectation values, $\langle|\varphi|^2\rangle_0$  and $\langle T^\mu_\nu\rangle_0$, and finite temperature contributions from the particles and antiparticles, $\langle|\varphi|^2\rangle_T$ and $\langle T^\mu_\nu\rangle_T$. In the limit $T\rightarrow0$, the latter goes to zero. Both thermal densities are even periodic function of the magnetic flux with a period equal to the quantum flux and even function of the chemical potential. Besides, we have shown that these densities depend only on the ratio of the magnetic flux by the quantum one, $\alpha_0$, which is a Aharonov-Bohm-like effect. As we have already mentioned the main objectives of this paper are the investigations of the thermal corrections of the field squared and the energy-momentum tensor. Due to the quasiperiodic condition on the $z$-axis, the momentum along this direction is quantized, and thanks to the Abel-Plana formula, we could decompose the densities to the part induced by the string and the other by the compactification. In the limit $L\rightarrow\infty$ the latter vanishes.

The total thermal correction to the field squared contains the contributions coming from Minkowski spacetime, Eq. \eqref{phi2_M_s}, plus the compactification in Minkowski spacetime, Eq. \eqref{phi2_Min_c}, and the part induced by the high dimensional compactified cosmic string. The latter is composed by pure cosmic string part, $\langle|\varphi|^2\rangle_{Ts}$,  Eq. \eqref{phi2s}, and the contribution induced by the compactification, $\langle|\varphi|^2\rangle_{Tc}$, Eq. \eqref{phi2c}. The main objective of this paper was to analyze the contribution induced by the compactified cosmic string spacetime. In this sense we have investigated in detail various asymptotic regime for both contributions. For massless field the results obtained are given in \eqref{phi2s0} and \eqref{phi2c0}. The behavior of these densities as function of temperature are given in \eqref{T_low} and \eqref{T_low1} for low temperature limit. To investigate the behavior of these densities in high temperature regime an alternative expression, convenient for this limit, was provided with the use of Eq. \eqref{Resum}. The results obtained are given by \eqref{T_high} and \eqref{T_high1}. As we can see, in low temperature regime both contributions go to zero exponentially, and in the opposite regime, i.e., in the high temperature regime, the leading terms increase linearly  with $T$. These behaviors are exhibited in the right plots of Figures  \ref{fig1} and \ref{fig2}, for $\langle|\varphi|^2\rangle_{Ts}$ and  $\langle|\varphi|^2\rangle_{Tc}$, respectively, considering $D=3$. Also in the right plot of Fig. \ref{fig1} we exhibit the behavior of  $\langle|\varphi|^2\rangle_{Ts}$ as function of the distance to the string's core, and the right plot of Fig. \ref{fig2}  the behavior of $\langle|\varphi|^2\rangle_{Tc}$ as function of the ration $r/L$. 

Another analysis developed in this paper was the thermal correction to the energy-momentum tensor induced by the compactified cosmic string, $\langle T^\mu_\nu\rangle_T$. Similarly to what happened in the analysis of the field squared,  we decomposed this correction in a contribution induced by the cosmic string without compactification, Eq. \eqref{EM-cosmic-string}, and induced by the compactification, Eq. \eqref{EM-compactification}, whose components are given in \eqref{G-functions}. Considering only the thermal energy density, $\langle T^0_0\rangle_T$, we could observe that it is finite at the string's core in absence of magnetic flux; however in presence of it the situation changes. This density is finite at $r=0$ for $q|\alpha_0|>1$ and diverges  as $1/(mr)^{2(1-q|\alpha_{0}|)}$ on the string's core for $q|\alpha_0|<1$. The plots in Fig. \ref{Fig3} exhibit this characteristic for $\langle T^0_0\rangle_{Ts}$, considering $D=3$.  The left panel with $q|\alpha_0|<1$, presents a divergence at $r=0$, while the right one, with $q|\alpha_0|>1$, is finite at the core. Moreover, these plots also display changes in the behavior of $\langle T^0_0\rangle_{Ts}$ for different ratios of $T/m$. As we can see the intensity of the thermal energy density increases with the temperature. This fact is reinforced by its asymptotic behavior for large $T$. The analysis of $\langle T^0_0\rangle_{Ts}$ and $\langle T^0_0\rangle_{Tc}$ in the regime of low temperature were expressed in Eq.s \eqref{EM_Tlow_s} and \eqref{EM_Tlow_c}, respectively. To analyze  the behavior of these densities in the high temperature regime, we adopted the same procedure as we did in the evaluation of the field squared, by using the identity \eqref{Resum}. After some intermediate steps, we obtained Eq. \eqref{EM-cs-High-Temp} for the correction induced by the string and Eq. \eqref{EM-c-High-Temp} for the correction induced by the compactification. The leading terms for both corrections increase linearly with temperature. Considering $D=3$, in Fig. \ref{Fig_T00cs_TperM} we exhibit the behavior of $\langle T^0_0\rangle_{Ts}$ as a function of the ratio $T/m$, for $q=2.5$, $|\tilde{\mu}|/m=0.5$, $mr=1$, $\xi=0$ and different values of $\alpha_{0}$. Moreover, in Fig. \ref{Fig_T00c_TperM}, we display the behavior of $\langle T^0_0\rangle_{Tc}$  for fixed $q=2.5$, $mr=1$, $L/m=1$, $\xi=0$, $|\tilde{\mu}|/m=0.5$, $\tilde{\eta}=0.5$ and different values of $\alpha_{0}$. Depending on the value of $\alpha_0$ the balance between particles and antiparticles may reverse the signs of both densities.

Finally, we would like to highlight the fact that the induced field squared and the energy-momentum tensor present a strong dependence with the temperature.  In fact, these densities  are amplified by thermal effects. We can say that this is one of the most important results presented in this paper, and that these results may have some application  in the early cosmology where the temperature of the Universe was really high.

\section*{Acknowledgment}
 W.O.S is supported under grant 2022/2008, Paraíba State Research Foundation (FAPESQ). E.R.B.M is partially supported by CNPq under Grant no 301.783/2019-3.

\end{document}